\documentclass[journal]{IEEEtran}
\usepackage{graphicx}
\graphicspath{{figures/},{pics}}
\usepackage{cite}
\usepackage{subfigure}
\usepackage{amsmath,amssymb,amsfonts}
\usepackage{algpseudocode}
\usepackage{algorithm}
\usepackage{setspace}
\usepackage{array}
\usepackage{makecell}

\usepackage{color,soul}

\ifCLASSINFOpdf
\else
\fi
\begin{document}
%
\title{Multi-frequency Electromagnetic Tomography for Acute Stroke Detection Using Frequency-\\Constrained Sparse Bayesian Learning }
%
%
%

\author{Jinxi~Xiang,~\IEEEmembership{Student Member,~IEEE,}
	Yonggui~Dong,
	and~Yunjie~Yang,~\IEEEmembership{Member,~IEEE}
	\thanks{This work was support in part by National Natural Science Foundation of China under grant 61671270. The work of J. Xiang was supported in part by the China Scholarship Council under Grant 201906210259. (Corresponding authors: Yunjie Yang and Yonggui Dong)}
	\thanks{Jinxi Xiang is with the Department of Precision Instrument, Tsinghua University, Beijing 100084, China, and also with the Agile Tomography Group, University of Edinburgh, Edinburgh, EH9 3FG, U.K. (xiangjx16@mails.tsinghua.edu.cn).}
	\thanks{Yonggui Dong is  with the Department of Precision Instrument, Tsinghua University, Beijing 100084, China.(dongyg@mail.tsinghua.edu.cn).}
	\thanks{Yunjie Yang is with the Agile Tomography Group, University of Edinburgh, Edinburgh, EH9 3FG, U.K..(y.yang@ed.ac.uk).}
}

%
%

\markboth{Journal of \LaTeX\ Class Files,~Vol.~14, No.~8, Jan.~2020}%
{J. Xiang \MakeLowercase{\textit{et al.}}: mfEMT for Acute Stroke Detection Using Frequency-Constrained Sparse Bayesian Learning}
%



\maketitle

\begin{abstract}
	Imaging the bio-impedance distribution of the brain can provide initial diagnosis of acute stroke. This paper presents a compact and non-radiative tomographic modality, i.e. multi-frequency Electromagnetic Tomography (mfEMT), for the initial diagnosis of acute stroke. The mfEMT system consists of 12 channels of gradiometer coils with adjustable sensitivity and excitation frequency.  To solve the image reconstruction problem of mfEMT, we propose an enhanced Frequency-Constrained Sparse Bayesian Learning (FC-SBL) to simultaneously reconstruct the conductivity distribution at all frequencies. Based on the Multiple Measurement Vector (MMV) model in the Sparse Bayesian Learning (SBL) framework, FC-SBL can recover the underlying distribution pattern of conductivity among multiple images by exploiting the frequency constraint information. 
	{A realistic 3D head model was established to simulate stroke detection scenarios, showing the capability of mfEMT to penetrate the highly resistive skull and improved image quality with FC-SBL.} 
	Both simulations and experiments showed that the proposed FC-SBL method is robust to noisy data for image reconstruction problems of mfEMT compared to the single measurement vector model, which is promising to detect acute strokes in the brain region with enhanced spatial resolution and in a {baseline-free} manner. 
\end{abstract}

%
\IEEEpeerreviewmaketitle

\section{Introduction}
%
%
%
%

\IEEEPARstart{S}{troke} is the second most common cause of death worldwide, and the third most common cause of disability \cite{feigin2017global}. There is a significant increase in stroke burden across the world, especially in developing countries. There are two types of strokes: ischaemic, and hemorrhagic and among them, around 80 out of 100 are ischaemic strokes \cite{van2015type}. It is now possible to treat acute stroke with thrombolytic drugs but it must be executed within 3-6 hours of stroke onset.  The single most important factor for saving lives and for successful patient recovery is the time from incidence to treatment. Brain imaging must be conducted before treatment, in order to differentiate these two strokes as the thrombolytic drugs would worsen the case of hemorrhagic stroke \cite{packham2012comparison}. Existing imaging techniques for stroke diagnosis include Positron Emission Tomography (PET), diffusion/perfusion-weighted Magnetic Resonance Imaging (MRI), and  Computed Tomography (CT) \cite{Choi2019MonitoringAS}. But on this occasion, their applications are restrained due to long diagnosis time and/or limited accessibility. A compact, fast, and cost-effective solution for early acute stroke detection is highly desirable. {The aim is to facilitate early diagnosis of these life-threatening conditions, preferably already before arrival to the hospital, thereby improving medical outcomes. }

In the recent decade, Electrical Impedance Tomography (EIT) has been investigated for acute stroke detection  through indirectly imaging the bio-impedance change induced by acute stroke  \cite{romsauerova2006multi, aristovich2014method , packham2012comparison, malone2014stroke}. However, the presence of highly resistive skull remains challenging as it may block the excitation current flowing through the head. In addition, skin-to-electrode contact impedance varies with surface conditions which is unpredictable \cite{zolgharni2009imaging}. 
{To address this problem, Jiang \textit{et al.}\cite{jiang2019capacitively} proposed capacitively coupled electrodes to improve the sensitivity to poor electrode contact. As an alternative, Ljungqvist \textit{et al.} in\mbox{\cite{ljungqvist2017clinical}} proposed Microwave Tomography (MWT) to detect strokes based on the propagation of microwaves in brain tissues, which can penetrate the skull with little attenuation. In this work, antennas need to be applied with a certain pressure to attach on the scalp, thereby compressing the hair and removing the air \cite{fhager20193d}. In practice, it would be desirable to remove direct contact with test subjects.}

Multi-frequency Electromagnetic Tomography (mfEMT) is a non-contact and non-invasive imaging technique \cite{xiang2019design}. Instead of applying excitation current through contact electrodes, mfEMT employs inductive coils to generate magnetic fields based on the eddy current effect. It can penetrate the highly resistive skull easily 
{and moreover, reduce modeling errors of electrode placement as the sensor placement is independent from head shape.} 
With mfEMT, bio-impedance can be measured by multi-frequency excitation spreading across the bandwidth of interest. There are two imaging modes of mfEMT, i.e. Time-Difference (TD) imaging and Frequency-Difference (FD) imaging. TD imaging requires a before-lesion data set and a measurement data set. For acute stroke imaging, the before-lesion data set cannot be easily obtained as patients present after the event. FD imaging measures multi-frequency data in a short time interval without requiring a reference data set, and uses the differences between selected frequencies for imaging, making it more promising for baseline-free acute stroke imaging.

Thus far, the potential application of mfEMT for intracranial hemorrhage detection has been preliminarily investigated in \cite{zolgharni2009imaging,  xiao2019threeD, Xiao2018Multif}. But these work is mainly based on analytical models and simulation data. There are still needs to (i) perform proof-of-concept validation on a feasible experimental platform, and (ii) develop high-resolution image reconstruction algorithms to effectively visualize  anomaly.

Motivated by this, in this paper, we first report a 12-channel mfEMT system and then propose a FD image reconstruction approach named Frequency-Constrained Sparse Bayesian Learning (FC-SBL) {based on the linearized mfEMT model} for {baseline-free},  anomaly detection under noisy scenarios. Sparse Bayesian Learning (SBL) has attracted attention in recent years for solving the inverse problem \cite{nissinen2007bayesian, Zhang2013ExtensionOS, Yang2018ArtificialNN, Liu2018ImageRI,Liu2019AcceleratedSS, liu2020efficient}. The fundamental idea of FC-SBL is to exploit the correlation among images under a set of excitation frequencies by extending the SBL framework. Mathematically, such correlations lead to a constrained optimization problem that promotes the signal's group-sparsity and its rank-deficiency. We design a frequency-constrained block sparsity prior that incorporates both the frequency and spatial correlation of conductivity distribution. The signal and noise statistics are learned directly from the data by SBL. We then demonstrate the image reconstruction improvement of the proposed method through both simulations and experiments.

The paper is organized as follows. In Section \ref{sec:mfEMT}, we present the sensing principle and measurement system of mfEMT. In Section \ref{sec:mfEMT_problem}, we formulate two fundamental problems of mfEMT, i.e. the forward and inverse problem. Then in Section \ref{sec:FC-SBL}, we present the FC-SBL algorithm to solve the inverse problem. Section \ref{sec:Experiment} gives experimental results and {Section \ref{sec:discussion} proposes some discussions.} Section \ref{sec:conclusion} concludes the paper.

\section{mfEMT system}
\label{sec:mfEMT}
\subsection{Sensing Principle}
mfEMT images the conductivity of an object by measuring the mutual inductance between coils placed around its periphery. The object is excited in a magnetic field (primary field)  produced by a current flowing in a coil. A resulting electrical field in the object is generated to induce eddy current. Consequently, a secondary magnetic field can be measured externally. Our previous work reported a gradiometer coil with significantly improved sensitivity \cite{xiang2019design} to measure the secondary field. The sensitivity of the gradiometer coil is governed by:
\begin{equation}
	s_{g}=\frac{\Delta \varphi_{g}}{\Delta \sigma}=-\frac{V_{0}\left(P_{1}-P_{2}\right) \omega \mu_{0}}{V_{\mathrm{res}}}
	\label{eq:sensitivity}
\end{equation} 
where $\Delta \varphi_{g}$ denotes the phase response of the secondary field caused by conductivity change $\Delta \sigma$. $V_0$ is the voltage caused by the primary magnetic field.  $P_1$ and $P_2$ are the geometrical factors concerning the size and shape of the object and its position relative to the coil. $\omega$ is the excitation frequency and $\mu_0$ is the permeability in free space. $V_{\mathrm{res}}$ is a key parameter of sensitivity which is tuned by the residual voltage of two differential receiver coils. 

\subsection{System Structure}

\begin{figure*}[tbp]
	\centering
	\includegraphics[width = 7in]{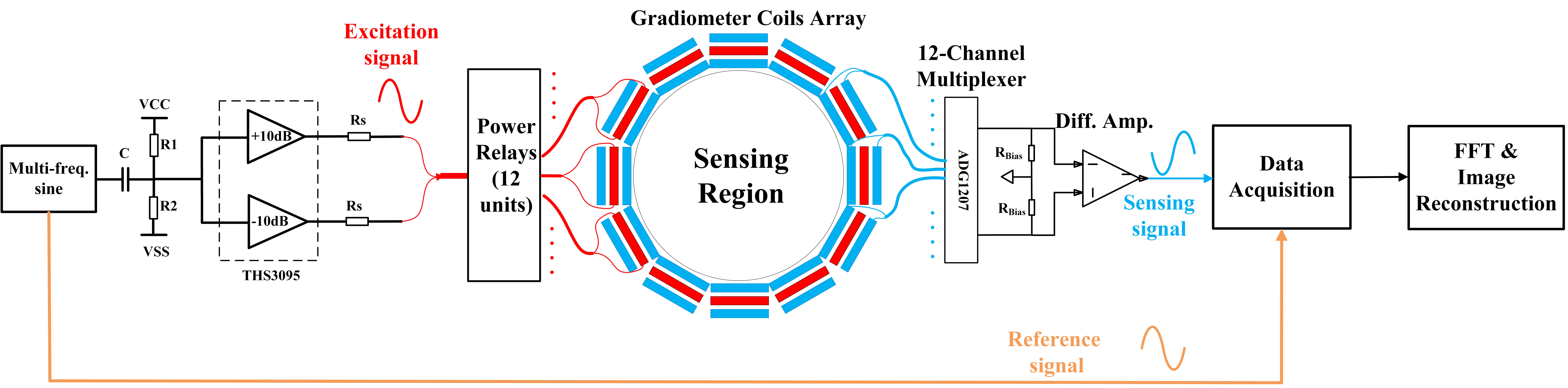}
	\caption{The 12-channel mfEMT system using gradiometer coils. {
			It comprises 4 modules:
			(1) sensor array;
			(2) excitation module;
			(3) front-end circuit and data acquisition modules;
			(4) phase demodulation and image reconstruction module.
		}}
		\label{fig:SystemRedPitaya}
	\end{figure*}
	
	The developed mfEMT system comprises 4 modules (see Fig.\ref{fig:SystemRedPitaya}):
	(1) a sensor array consisting of 12 gradiometer coils;
	(2) an excitation module to drive gradiometer coils;
	(3) the front-end circuit and data acquisition modules based on Red Pitaya, an open-source hardware platform with dual ADCs and DACs \cite{REDPITAYA};
	(4) phase demodulation by Fast Fourier Transform (FFT) and image reconstruction.
	
	The operation principle of the mfEMT system is as follows. First, one of the 12 excitation channels is enabled and the rest of the other 11 excitation channels are disabled as open circuits. Second, multi-frequency sine waves are generated by Red Pitaya. Then, the 12 differential sensing coils are sequentially selected and the sensing signals are multiplexed to be acquired.  One frame of data covering all the excitation/sensing coil combinations consists of 144 (12$\times$12) measurements.
	
	
	\section{Forward and Inverse Problem of mfEMT}
	\label{sec:mfEMT_problem}
	\subsection{Forward Problem}
	
	Two problems of mfEMT need to be solved, i.e. the forward problem and inverse problem. The forward problem is to determine measurements (phase values) given the conductivity distribution \cite{ma2017magnetic}, as expressed by:
	\begin{equation}
		\boldsymbol{\varphi} =\mathbf{F}(\boldsymbol{\sigma})+\mathbf{v}
		\label{eq:forward1}
	\end{equation}
	where $\boldsymbol{\sigma} \in \mathbb{R}^{M}$ is the conductivity distribution; $\boldsymbol{\varphi} \in \mathbb{R}^{N}$ represents the noisy measurements, and $N\ll M$; $\mathbf{v}$ is the noise vector; $\mathbf{F}$ is a nonlinear function mapping the conductivity distribution to measurements.
	{A commonly used, simplified linear model is adopted in this work}:
	\begin{equation}
		\Delta \boldsymbol{\varphi} \approx  \frac{\partial \mathbf{F}(\boldsymbol{\sigma})}{\partial \boldsymbol{\sigma}}\Delta \boldsymbol{\sigma} + \mathbf{v} =  \mathbf{J} \cdot \Delta \boldsymbol{\sigma}+ \mathbf{v}
		\label{eq:smatrix}
	\end{equation}
	
	The sensitivity matrix $\mathbf{J}\in \mathbb{R}^{N \times M}$ maps the conductivity distribution to measurements, which is solved by \cite{rosell2001sensitivity}:
	\begin{equation}
		\mathbf{J}(\Omega) = \frac{k(\omega)}{I_1 I_2}  \mathbf{B_1} \cdot \mathbf{B_2} 
		\label{eq:magnetic_field_dot}
	\end{equation}
	where $\Omega$ is the spatial coordinates; $\mathbf{B_1}$ is the magnetic field produced by a current $I_1$ injected into the excitation coil; $\mathbf{B_2}$ is the magnetic field produced by a current $I_2$ injected into the differential sensing coil; {$k(\omega)$ is a scalar given a known working frequency $\omega$. The forward problem of mfEMT is solved with a numerical model in our previous work \cite{xiang2019design}.}
	
	\subsection{Inverse Problem}
	Image reconstruction of mfEMT is a typical inverse problem. The objective is to estimate $\boldsymbol{\sigma}$ from $\boldsymbol{\varphi}$. A general optimization framework can be formulated as: 
	\begin{equation}
		\widehat{\boldsymbol{\sigma}}=\underset{\boldsymbol{\sigma} \in \mathbb{R}^{M}}{\arg \min }\{ d(\boldsymbol{\sigma})+r(\boldsymbol{\sigma}) \} 
		\label{eq:opt}
	\end{equation}
	where $d(\boldsymbol{\sigma})$ is the data-fidelity term that penalizes the mismatch to the measurements and $r(\boldsymbol{\sigma})$ is the regularizer that imposes a prior information of the conductivity. Two common regularizers include the spatial sparsity-promoting penalty $r(\boldsymbol{\sigma}) \triangleq\|\boldsymbol{\sigma}\|_{\ell_{1}}$ and Total Variation penalty ${r}(\boldsymbol{\sigma}) \triangleq\|\mathbf{D} \boldsymbol{\sigma}\|_{\ell_{1}}$, where $\mathbf{D}$ is a discrete gradient operator \cite{baraniuk2007compressive}. 
	
	In this work, we turn to a statistical perspective on sparsity-promoting regularization. Optimization problem of (\ref{eq:opt}) is interpreted in a Bayes perspective within the SBL framework to maximize the posterior $p(\boldsymbol{\sigma}| \boldsymbol{\varphi})$:
	\begin{equation}
		\underset{\boldsymbol{\sigma}}{\arg \max } \ p(\boldsymbol{\sigma} | \boldsymbol{\varphi}) \triangleq \underset{\boldsymbol{\sigma}}{\arg \min }\left[-\log p(\boldsymbol{\varphi} | \boldsymbol{\sigma})-\lambda\log p (\boldsymbol{\sigma})\right]
		\label{eq:map}
	\end{equation} 
	where, the data-fidelity term $d(\boldsymbol{\sigma}) = -\log p(\boldsymbol{\varphi}| \boldsymbol{\sigma})$ models the data likelihood. This term encapsulates the physics model for generation of measurement $\boldsymbol{\varphi}$. $r(\boldsymbol{\sigma}) = -\lambda\log p (\boldsymbol{\sigma})$ is the regularization term  by using prior $p (\boldsymbol{\sigma})$ that is elaborately  handcrafted or learned in accordance with the prior of $\boldsymbol{\sigma}$. 
	
	{Sparse Bayesian learning is a statistical framework, which assigns probabilities rather than deterministic values to model parameters by combining a data model with a prior model \cite{dashti2013bayesian}. The posterior probability of the model parameter conditioned on observed data describes all possible solutions to the inverse problem along with their probabilities, and it is essential for uncertainty quantification \cite{adler2018deep}. Compared to the conventional approaches, the capability of quantifying the uncertainty makes it robust to data disturbance \cite{Liu2018ImageRI}.}

	\section{Image Reconstruction with FC-SBL}
	\label{sec:FC-SBL}
	
	\subsection{Multi-Frequency Measurement Preprocessing}
	\label{sec:preprocess}
	
	Given that we acquire a sequence of multi-frequency measurements $\left\{ \boldsymbol{\varphi}_{f_0},\boldsymbol{\varphi}_{f_1}, \ldots, \boldsymbol{\varphi}_{f_L} \right\}$ of mfEMT. The measurement at the lowest frequency $\boldsymbol{\varphi}_{f_0}$ is selected as a reference. Generally, both the target object (acute stroke) and the background (other matters in the brain) are frequency-dependent. Thus, simple frequency difference $\Delta \boldsymbol{\varphi}_{f_i} = \boldsymbol{\varphi}_{f_i}-\boldsymbol{\varphi}_{f_0}$ cannot cancel out the background signal. 
	{Assume there exists anomalies in the sensing region with a homogeneous background, we first apply background subtraction to counteract the background change and enhance the local contrast of the anomalies. The phase change due to the frequency-dependent anomaly can be differentiated by decomposing $ \boldsymbol{\varphi}_{f_i} $:} 
	\begin{equation}
		\boldsymbol{\varphi}_{f_i} = \underbrace{ \boldsymbol{\varphi}_{f_0}+ \Delta \boldsymbol{\varphi}_{bg} }_\text{background signal} + \underbrace{\Delta \boldsymbol{\varphi}_{ano} }_\text{target signal}
		\label{eq:weighted}
	\end{equation}
	where $\Delta \boldsymbol{\varphi}_{bg}$ is the phase response induced by the background, whereas $\Delta \boldsymbol{\varphi}_{ano}$  by the target anomaly. Since the background is assumed to be homogeneously conductive, it could be represented by $\Delta \boldsymbol{\varphi}_{bg} = \alpha_i \mathbf{J}\boldsymbol{1}\in \mathbb{R}^{N\times1}$. Therefore, the variance induced by the target is
	\begin{equation}
		\Delta \boldsymbol{\varphi}_{obj} = ( \boldsymbol{\varphi}_{f_i}-  \boldsymbol{\varphi}_{f_0}) - \alpha_i \mathbf{J}\boldsymbol{1}
		\label{eq:weighted_diff}
	\end{equation}
	where $\alpha_i =\frac{\left\langle ( \boldsymbol{\varphi}_{f_i}-  \boldsymbol{\varphi}_{f_0}), \mathbf{J}\boldsymbol{1}\right\rangle}{\left\langle\mathbf{J}\boldsymbol{1}, \mathbf{J}\boldsymbol{1}\right\rangle}$ is a projection coefficient; $\left\langle , \right\rangle$ denotes the inner product. In consequence, the weighted  differences of multiple measurements are:
	\begin{equation}
		\left\{
		\begin{array}{lr}
			\Delta \boldsymbol{\varphi}_{f_1} & = ( \boldsymbol{\varphi}_{f_1}-  \boldsymbol{\varphi}_{f_0}) - \alpha_1 \mathbf{J}\boldsymbol{1}\\
			\Delta \boldsymbol{\varphi}_{f_2} & = ( \boldsymbol{\varphi}_{f_2}-  \boldsymbol{\varphi}_{f_0}) - \alpha_2 \mathbf{J}\boldsymbol{1} \\
			\vdots &  \\
			\Delta \boldsymbol{\varphi}_{f_L} & = ( \boldsymbol{\varphi}_{f_L}-  \boldsymbol{\varphi}_{f_0}) - \alpha_L \mathbf{J}\boldsymbol{1}
		\end{array}
		\right.
		\label{eq:weighted_FD}
	\end{equation}

	As revealed in  (\ref{eq:sensitivity}), the sensitivity of the gradiometer coil is linearly proportional to the excitation frequency. To use one identical sensitivity matrix $\mathbf{J}$ throughout the algorithm development, the measurements are further normalized in accordance with frequency:
	\begin{equation}
		\Delta \boldsymbol{\varphi}_{\text{norm}} = \left[ \Delta \boldsymbol{\varphi}_{f_1},\ldots, \Delta \boldsymbol{\varphi}_{f_L}  \right] 
		\left[
		\begin{array}{ccc}
			f_1 & & \mathbf{0} \\ 
			{} & {\ddots} &\\ 
			{\mathbf{0}} & & f_L
		\end{array}
		\right]^{-1}
		\label{eq:phase_n}
	\end{equation}
	
	Thereafter, we use $\left \{ \mathbf{y}_i\right \}, i=1\cdots,L$ to denote $\Delta \boldsymbol{\varphi}_{\text{norm}}$.
	
	\begin{figure*}[tbp]
		\centering
		\subfigure[{Frequency constraints: conductivity distribution patterns are identical whilst conductivity values increase with frequency. Frequency constraints among pixels are presented in rows after vectorization.}]{\includegraphics[height = 1.5 in]{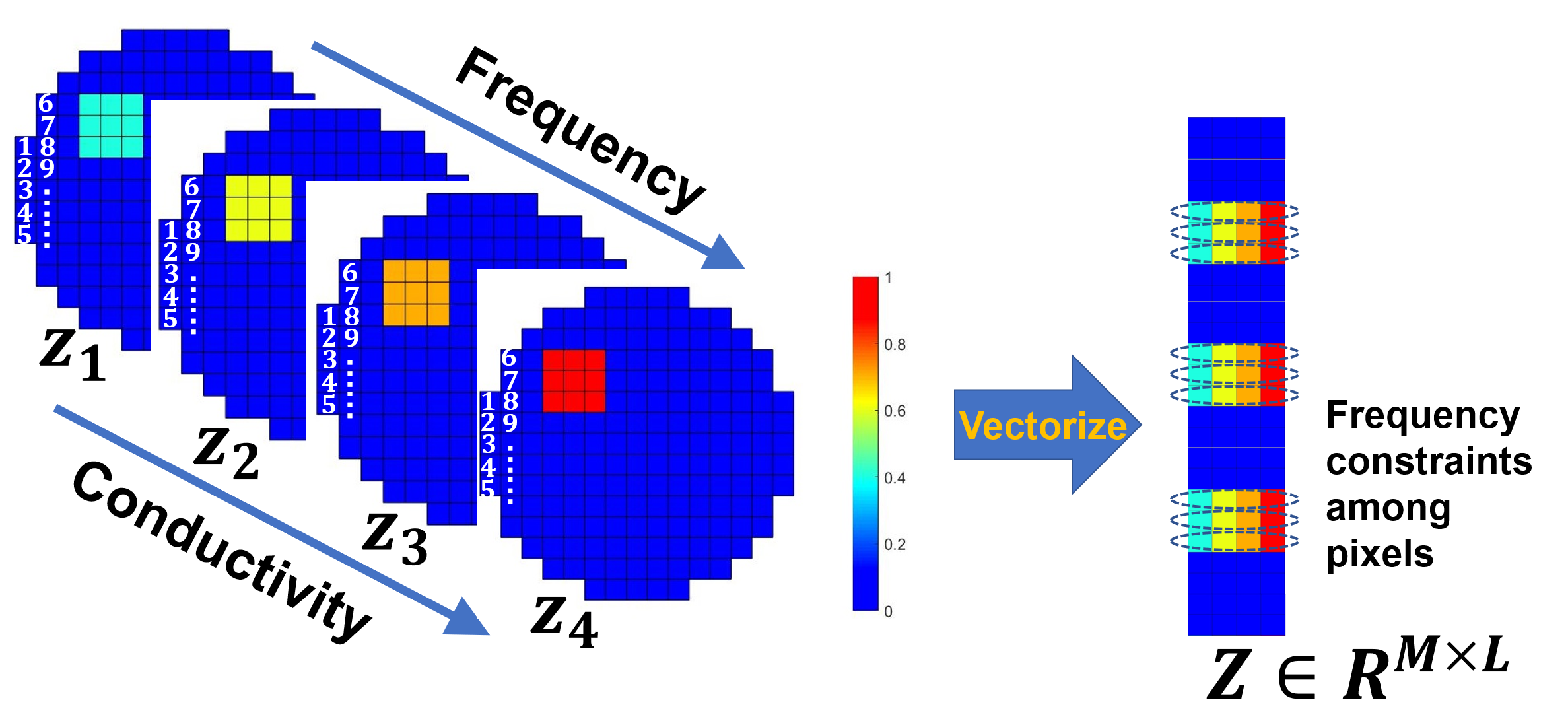}}
		\hspace{0.2in}
		\subfigure[{Multi-frequency measurement model in (\ref{eq:MMV3}): a 2D block partition structure $\mathbf{E}$ is embedded in the sensitivity matrix $\mathbf{J}$, generating a new matrix $\mathbf{\Phi}$. After block embedding, the unknown matrix $\mathbf{X}$ is block sparse, clustering the related elements (both frequency and spatial correlated) into one block.} ]{\includegraphics[height = 1.5 in]{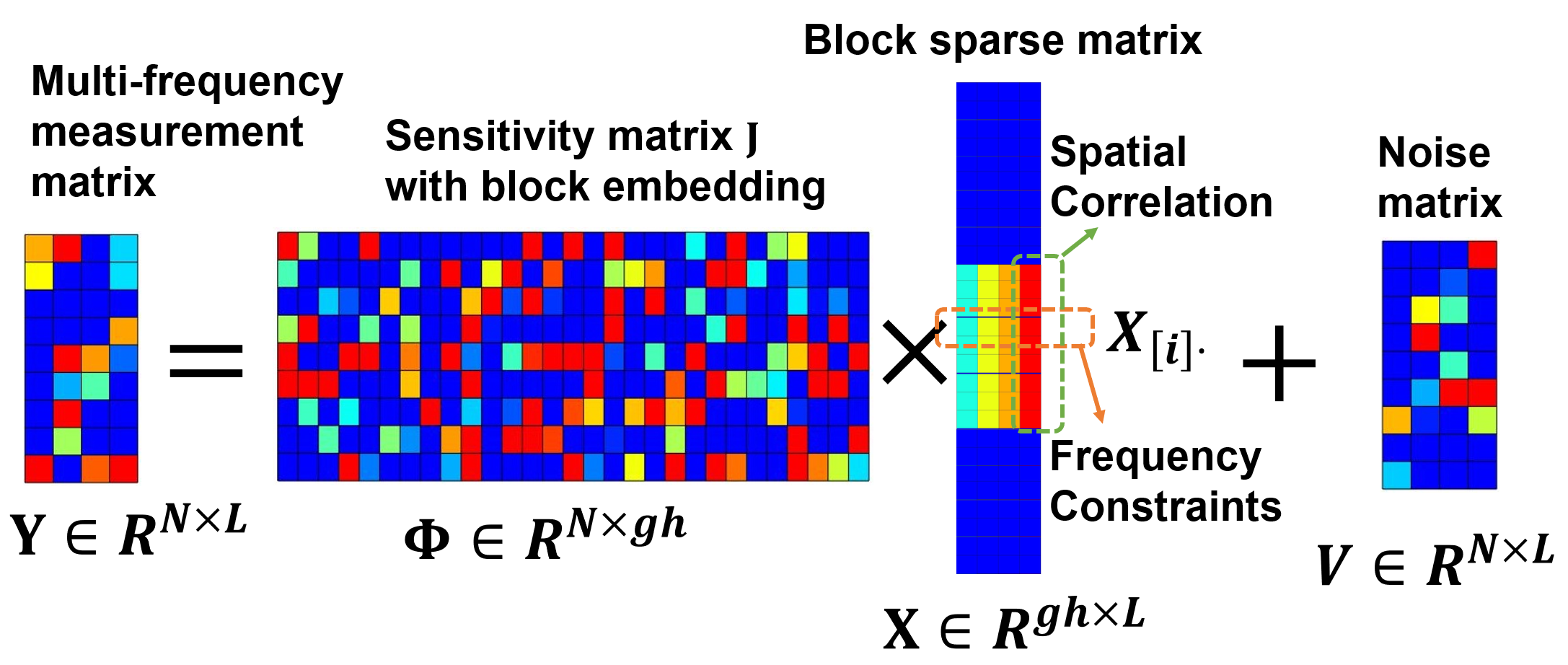}}
		\caption{{Schematic illustration of FC-SBL that exploits both spatial correlation and frequency constraints of the conductivity distribution. The figures show one case when the embedding block size is $h=9$. Any realistic conductivity distribution patterns can be represented by multiple overlapped blocks.}}
		\label{fig:FC-SBL}
	\end{figure*}
	
	\subsection{FC-SBL}
	\label{sec:FCSBL_body}
	In mfEMT,  multi-frequency measurements are acquired and then preprocessed as described in Section \ref{sec:preprocess}. As described in (\ref{eq:smatrix}), each measurement is governed by the same linearized equation:
	\begin{equation}
		\left\{
		\begin{array}{lr}
			\mathbf{y}_1 & =\mathbf{J} \mathbf{z}_1+\mathbf{v}_1 \\
			\mathbf{y}_2 & =\mathbf{J} \mathbf{z}_2+\mathbf{v}_2 \\
			\vdots &  \\
			\mathbf{y}_L & =\mathbf{J} \mathbf{z}_L+\mathbf{v}_L
		\end{array}
		\right.
		\label{eq:SMV}
	\end{equation}
	where $\{\mathbf{y}_1, \mathbf{y}_2, \dots, \mathbf{y}_L\}$, $\mathbf{y}_i \in \mathbb{R}^{N\times 1}$ ($i = 1,2,\dots L$) denotes the preprocessed measurements;  $\mathbf{J} \in \mathbb{R}^{N \times M}(N \ll M)$ represents the sensitivity matrix;   $\{\mathbf{z}_1, \mathbf{z}_2, \dots, \mathbf{z}_L\}$, $\mathbf{z}_i =\Delta \boldsymbol{\sigma_i} \in \mathbb{R}^{M\times 1}$ ($i = 1,2,\dots L$)  are the relative conductivity images to be solved; $\{\mathbf{v}_1, \mathbf{v}_2, \dots, \mathbf{v}_L\}$, $\mathbf{v}_i \in \mathbb{R}^{N\times 1}$ ($i = 1,\dots L$) are noise vectors.
	
	{As illustrated in Fig. \ref{fig:FC-SBL} (a), the conductivity values change over frequency but the distribution patterns are identical. It would be beneficial to take advantage of frequency constraints on distribution patterns if we organize the measurements into one matrix.} To achieve this goal, we extend  (\ref{eq:SMV}) to  the MMV model \cite{ziniel2012efficient}:
	\begin{equation}
		\mathbf{Y}=\mathbf{J}\mathbf{Z}+\mathbf{V}
		\label{eq:MMV1}
	\end{equation}
	where $\mathbf{Y} = [ \mathbf{y}_1, \mathbf{y}_2, \dots \mathbf{y}_L  ] \in \mathbb{R}^{N\times L}$, $\mathbf{Z} = [ \mathbf{z}_1, \mathbf{z}_2, \dots \mathbf{z}_L  ] \in \mathbb{R}^{M\times L}$, $\mathbf{V} = [ \mathbf{v}_1, \mathbf{v}_2, \dots \mathbf{v}_L  ]\in \mathbb{R}^{N\times L}$.  Each column of $\mathbf{Z}$ corresponds to a {vectorized conductivity image} at a given frequency and each row represents a pixel in the image.  {In practice,  the block partition of $\mathbf{Z}$ is unknown, meaning that the distribution could be arbitrary and there are no additional structures information of $\mathbf{Z}$ such as block/group structure \cite{zhang2013extension}.}
	
	Without requiring any prior knowledge of the block partition in $\mathbf{Z}$, we consider a general case that 2D square blocks with equal size of $h$ overlap with each other. This block structure is different from the 1D block in previous literatures \cite{Zhang2013ExtensionOS, Liu2018ImageRI}. Note that  the resulting algorithms are not very sensitive to the block structure (1D or 2D) since real partition can be learned during the SBL process. Empirical evidence showed that algorithmic performance can be further improved with 2D blocks because spatial correlations are inherently two-dimensional rather than one-dimensional. A 2D block partition structure  is embedded with a matrix $\mathbf{E}$:
	\begin{equation}
		\mathbf{z}_j \triangleq \mathbf{E} \mathbf{x_{\cdot j}} \triangleq\left[\mathbf{E}_{1}, \ldots,\mathbf{E}_i, \ldots \mathbf{E}_{g}\right]\left[\mathbf{x}_{1j}^{\mathrm{T}}, \ldots, \mathbf{x}_{ij}^{\mathrm{T}},\ldots, \mathbf{x}_{gj}^{\mathrm{T}}\right]^{\mathrm{T}}
		\label{eq:MMV2}
	\end{equation}
	where $j=1,\cdots, L$; $g\approx M$ is the total number of possible blocks in $\mathbf{z}_j$.  
	$\forall i=1, 2, \dots g$, $\mathbf{E}_{i} \triangleq\left[ \mathbf{e}_{b_1},\cdots, \mathbf{e}_{b_h}  \right]\in \mathbb{R}^{M \times h}$ is the $i$-th block structure;  $\mathbf{e}_{b_k}=\left[ 0, \cdots, 0,1,0, \cdots, 0\right]^T$ is a base vector where the $1$ appears at the $b_k$-th position, $k=1,2\cdots,h$. $\left\{ b_k \right\}$ is a set recording the pixel index of each element in the $i$-th block. 
	$\mathbf{x}_{ij}= \left[x_{(ih-h+1)j}, \ldots, x_{(ih)j}\right]^{\mathrm{T}} \in \mathbb{R}^{h \times 1}$ denotes the weights of each element in the block.

	Then, the spatial block-sparse underlying model in  (\ref{eq:MMV1}) is further expressed as (see Fig. \ref{fig:FC-SBL}):
	\begin{equation}
		\mathbf{Y}=\mathbf{J}\mathbf{Z}+\mathbf{V}=\boldsymbol{\Phi}\mathbf{X} + \mathbf{V}
		\label{eq:MMV3}
	\end{equation}
	where $\boldsymbol{\Phi} = \mathbf{J}\mathbf{E}$;  $\mathbf{X}= \left[ \mathbf{x_{\cdot 1}},\cdots \mathbf{x_{\cdot j}} \cdots \mathbf{x_{\cdot L}} \right]\in \mathbb{R}^{gh\times L} $. Note from Fig. \ref{fig:FC-SBL} that $\mathbf{X}$ has both inter-row (spatial block) and inter-column (frequency constraints)  correlation. We can formulate this structure into a block matrix form:
	\begin{equation}
		\mathbf{X} = \left[\begin{array}{c}{\mathbf{X}_{[1]\cdot}}^\mathrm{T},\ {\mathbf{X}_{[2]\cdot}}^\mathrm{T},\  {\ldots}, {\mathbf{X}_{[g]\cdot}}^\mathrm{T}\end{array}\right]^\mathrm{T}
		\label{eq:intrablock}
	\end{equation}
	$ \forall i=1,\cdots,g, \mathbf{X}_{[i]\cdot}\in \mathbb{R}^{h\times L} $ denotes the $i$-th block of all the column. It incorporates the spatial and frequency information of multiple measurements into each block matrix. This is the core idea of the FC-SBL method. 
	
	To solve $\mathbf{X}$ in  (\ref{eq:MMV3}), we reformulate the MMV model to frequency-constrained block SMV model \cite{Zhang2011SparseSR}:
	\begin{equation}
		\mathbf{y}_F = \mathbf{D} \mathbf{x}_F + \mathbf{v}_F 
		\label{eq:MMV4}
	\end{equation}
	by letting $\mathbf{y}_F = \operatorname{vec}\left(\mathbf{Y}^{T}\right)\in \mathbb{R}^{N L \times 1}$, $\mathbf{D} = \boldsymbol{\Phi} \otimes \mathbf{I}_{L}\in \mathbb{R}^{NL\times ghL}$, $\mathbf{x}_F = \operatorname{vec}\left(\mathbf{X}^{T}\right)\in \mathbb{R}^{ghL\times 1}$, $\mathbf{v}_F = \operatorname{vec}\left(\mathbf{V}^{T}\right)\in \mathbb{R}^{NL \times 1}$.
	$\otimes$ represents the Kronecker product of the two matrices. $\operatorname{vec}(.)$ denotes the vectorization of the matrix  by stacking its columns into a single column vector.
	
	To exploit the spatial block correlation and frequency-constrained information simultaneously, we design a covariance matrix $\mathbf{\Sigma_{0}}$ as a prior of the weights $\mathbf{x}_F$ using a zero-mean Gaussian distribution
	\begin{equation}
		p\left(\mathbf{x}_F ; \left\{\gamma_{i}, \mathbf{C}_{i}\right\}_{i=1}^{g}\right) \sim \mathcal{N}_{\mathbf{x}_F} \left(\mathbf{0}, \mathbf{\Sigma_{0}}\right)
		\label{eq:sbl1}
	\end{equation}
	where $\mathbf{\Sigma_0}$ is 
	\begin{equation}
		\mathbf{\Sigma}_{0}=\left[\begin{array}{ccc}{\gamma_{1} \mathbf{C}_{1}} & & {\mathbf{0}} \\ {} & {\ddots} & {} \\ {\mathbf{0}} & {} & {\gamma_{g} \mathbf{C}_{g}}\end{array}\right] \in \mathbb{R}^{g h L\times g h L}
		\label{eq:sbl2}
	\end{equation}
	$\mathbf{\Sigma_{0}}$ is a block diagonal matrix. $\gamma_i\geq 0(i=1,\ldots,g)$, most of which are zeros, determines the sparsity pattern of $\mathbf{X}$.  $\mathbf{C}_i\in \mathbb{R}^{hL\times hL}(i=1,\ldots,g)$ is a covariance matrix. 
	
	It is suggested that for many applications, the intra-block elements can be sufficiently represented by a first-order Auto-Regressive (AR) process to  model the correlation \cite{zhang2013extension, lin2014covariance} and thus, the corresponding covariance matrix is a Toeplitz matrix. Since  the covariance matrix $\left\{ \mathbf{C}_i\right\}_{i=1}^{g}$ incorporate the spatial and frequency information together, we model it with two different AR processes, i.e., $\left\{ \mathbf{A}_i \right\}_{i=1}^{g} = \text { Toeplitz }\left(\left[1, r_{si}, \dots, r_{si}^{h-1}\right]\right)$ for spatial correlation;  $\mathbf{B} =\text { Toeplitz }\left([1, r_f, \dots, r_f^{L-1}]\right)$ for frequency constraints. Thus,  $\mathbf{C}_i$ is regularized as the Kronecker product of two Toeplitz matrices (not commutative):
	\begin{equation}
		\mathbf{C}_i  = \mathbf{A}_i \otimes \mathbf{B}
		\label{eq:toep_times}
	\end{equation}
	
	In this way, the resulting prior $\mathbf{\Sigma}_0$  is written as:
	\begin{equation}
		\mathbf{\Sigma}_{0}=\left[\begin{array}{ccc}{\gamma_{1} \mathbf{A}_{1}} & {} &{\mathbf{0}} \\ {} & {\ddots} & {}\\ {\mathbf{0}} &{} &{\gamma_{g} \mathbf{A}_{g}}\end{array}\right] \otimes \mathbf{B} = \boldsymbol{\Pi}\otimes \mathbf{B}
		\label{eq:toep_otimes}
	\end{equation}

	Moreover, we assume that the elements in the noise vector $\left\{ \mathbf{v}_i\right\}$ are independent and each follows Gaussian distribution i.e.  $p\left(\mathbf{v}_i\right) \sim \mathcal{N}_{\mathbf{v}_i}(0, \lambda\mathbf{I})$. Considering the frequency constraints $\mathbf{B}$ in $\mathbf{v}_F$,  we get $p\left(\mathbf{v}_F\right) \sim \mathcal{N}_{\mathbf{v}_F}(0, \lambda\mathbf{I}\otimes \mathbf{B})$.
	
	{To solve (\ref{eq:MMV4}), we write down the posterior $p(\mathbf{x}_F|\mathbf{y}_F)$ using Bayesian rule by imposing the prior in (\ref{eq:toep_otimes}).}
	\begin{equation}
		p(\mathbf{x}_F | \mathbf{y}_F ; \lambda,\left\{\gamma_{i}, \mathbf{A}_{i}\right\}_{i=1}^{g}, \mathbf{B}) \sim \mathcal{N}_{\mathbf{x}_F|\mathbf{y}_F}\left(\boldsymbol{\mu}_{x}, \mathbf{\Sigma}_{x}\right)
		\label{eq:sbl3}
	\end{equation}
	where $\boldsymbol{\mu_x}$, $\mathbf{\Sigma}_{x}  $ are given by:
	\begin{equation}
		\begin{aligned}
			\boldsymbol{\mu}_x &=\boldsymbol{\Sigma }\mathbf{D}^{T}(\lambda \mathbf{I} \otimes \mathbf{B})^{-1} \mathbf{y}_{F} \\
			&=\operatorname{vec}\left(\mathbf{Y}^{T}\left(\lambda \mathbf{I}+\boldsymbol{\Phi} \boldsymbol{\Pi} \boldsymbol{\Phi}^{T}\right)^{-1} \mathbf{\Phi} \boldsymbol{\Pi}\right)
		\end{aligned}
		\label{eq:mux}
	\end{equation}
	\begin{equation}
		\begin{aligned}
			\boldsymbol{\Sigma}_x &=\left((\boldsymbol{\Pi} \otimes \mathbf{B})^{-1}+\mathbf{D}^{T}(\lambda \mathbf{I} \otimes \mathbf{B})^{-1} \mathbf{D}\right)^{-1} \\
			&=\left[\boldsymbol{\Pi}-\boldsymbol{\Pi} \boldsymbol{\Phi}^{T}\left(\lambda \mathbf{I}+\boldsymbol{\Phi} \boldsymbol{\Pi} \mathbf{\Phi}^{T}\right)^{-1} \boldsymbol{\Phi} \boldsymbol{\Pi}\right] \otimes \mathbf{B}
		\end{aligned}
		\label{eq:sigmax}
	\end{equation}

	The final estimation of conductivity is:
	\begin{equation}
		{\hat{\mathbf{Z}}}  = \mathbf{E} \hat{\boldsymbol{\mu}}_x = \mathbf{E} \boldsymbol{\Pi} \boldsymbol{\Phi}^{T}\left(\lambda \mathbf{I}+\boldsymbol{\Phi} \boldsymbol{\Pi} \boldsymbol{\Phi}^{\mathrm{T}}\right)^{-1} \mathbf{Y}
		\label{eq:mu_est}
	\end{equation}
	where $\hat{\boldsymbol{\mu}}_x$ is obtained by reshaping $\boldsymbol{\mu}_x$ to $\mathbb{R}^{gh\times L}$ dimension. The estimation of hyperparameter $\Theta = \left\{ \lambda, \left\{\gamma_{i}, \mathbf{A}_i \right\}_{i=1}^{g}, \mathbf{B} \right\}$ is the main body of FC-SBL.
	
	\renewcommand{\algorithmicrequire}{\textbf{Input:}}
	\renewcommand{\algorithmicensure}{\textbf{Output:}}
	\begin{algorithm}
		\caption{: Frequency-Constrained Sparse Bayesian Learning (FC-SBL) for mfEMT Image Reconstruction}\label{euclid}
		\begin{algorithmic}[1]
			\Require {$\left\{\mathbf{Y}, \mathbf{J}, h, \epsilon_{\mathrm{min}}, \vartheta_{\mathrm{max}} \right\}$} 
			\Comment{multi-frequency measurement vectors $\mathbf{Y}\in \mathbb{R}^{N\times L}$; sensitivity matrix of mfEMT $\mathbf{J}\in \mathbb{R}^{N\times M}$; spatial block size $h$; the minimum error bound $\epsilon_{\mathrm{min}}$; and the  maximum iteration steps $\vartheta_{\mathrm{max}}$}.
			\Ensure {$\hat{\mathbf{Z}} = \left\{ \hat{\mathbf{z}}_1,\ldots, \hat{\mathbf{z}}_L  \right\}$}
			\Comment{reconstructed conductivity distribution of $L$ frequencies}
			\State \textbf{Initialize: }{$$\begin{array}{l}{ 
					\epsilon=1, \vartheta=0, \boldsymbol{\mu}_{x}=\mathbf{0}_{g h L\times L}, \mathbf{\Sigma}_{x}=\mathbf{0}_{g h L \times g h L}} \\ 
				{\gamma_i = 1, i=1,\cdots,g} \\ 
				{\lambda=0.01 \times \frac{1}{L-1} \sum_{i=1}^{L} \left( \sqrt{\frac{1}{M-1} \sum_{j=1}^{M}\left|y_{ji}-\overline{\mathbf{y}}_i \right|^{2}}  \right) } \\ 
				{\mathbf{A}_{i}= \text { Toeplitz }\left(\left[0.9^{0}, \ldots, 0.9^{h-1}\right]\right),i=1,\cdots,g }\\ 
				{ \mathbf{B} = \text { Toeplitz }\left(\left[0.9^{0}, \ldots, 0.9^{L-1}\right]\right)}
				\end{array}$$}
			\While{ $\epsilon > \epsilon_{\mathrm{min}}$ and $ \vartheta < \vartheta_{\mathrm{max}}$}
			\State Update $\boldsymbol{\mu}_x$ using (\ref{eq:mux});
			\State Upadate $\boldsymbol{\Sigma_x}$ using  (\ref{eq:sigmax}); 
			\State Update $\left\{ \gamma_i \right\}_{i=1}^{g}$ using  (\ref{eq:lr_gamma});
			\State Update ${\mathbf{B}}$ using (\ref{eq:lr_Bplus}) (\ref{eq:lr_B});
			\State Update  $r_f$ with (\ref{eq:toep_rf}); regularize $\mathbf{B}$ in Toeplitz matrix;
			\State Update $\mathbf{A}_i$ using (\ref{eq:lr_Ai});
			\State Update  $r_{si}$ with (\ref{eq:toep_rs}); regularize $\mathbf{A}_i$ in Toeplitz matrix;
			\State Update $\lambda$ using (\ref{eq:lr_lambda});
			\State Estimate  $ \epsilon=\left\|\boldsymbol{\mu}_{x}^{(\vartheta)}-\boldsymbol{\mu}_{x}^{(\vartheta-1)}\right\|_{2} /\left\|\boldsymbol{\mu}_{x}^{(\vartheta)}\right\|_{2}$;
			\State Iteration update $\vartheta \leftarrow \vartheta +1$.
			\EndWhile\label{euclidendwhile}
			\State \textbf{return} $\hat{\mathbf{Z}} = \mathbf{E}\hat{\boldsymbol{\mu}}_x\qquad$   \Comment{$\mathbf{E}\in \mathbb{R}^{M\times ghL}$ is a predefined matrix of embedding block structure.}
		\end{algorithmic}
	\end{algorithm}
	
	\subsection{Hyper Parameter Estimation}
	
	The hyper parameters $\Theta = \left\{ \lambda, \left\{\gamma_{i}, \mathbf{A}_i \right\}_{i=1}^{g}, \mathbf{B} \right\}$ can be estimated by a Type II maximum likelihood procedure \cite{Tipping2003BayesianIA}, yielding the effective cost function:
	\begin{equation}
		\mathcal{L}(\Theta)=\mathbf{\mathbf{y}_F}^{\mathrm{T}} \boldsymbol{\Sigma}_{\mathbf{y}_F}^{-1} \mathbf{\mathbf{y}_F}+\log \left|\boldsymbol{\Sigma}_{\mathbf{y}_F}\right|
		\label{eq:emcost}
	\end{equation}
	where $\mathbf{\Sigma}_{\mathbf{y}}=\lambda \mathbf{I} \otimes \mathbf{B}+\mathbf{D}(\mathbf{\Pi} \otimes \mathbf{B}) \mathbf{D}^{T}$.

	By optimizing the cost function with respect to each parameter in $\Theta$, we derive the following learning rules. {First, we derive the learning rule for $\gamma_i$ using a bound-optimization method \cite{Zhang2011SparseSR}, considering an upper-bound for the second term in (\ref{eq:emcost}), and then minimize the upper-bound of the cost function.}
	\begin{equation}
		\gamma_{i} \leftarrow \sqrt{\frac{ \operatorname{Tr}\left(\mathbf{X}_{[i]\cdot}  \mathbf{B}^{-1} \mathbf{X}_{[i]\cdot}^{\mathrm{T}} \mathbf{A}_{i}^{-1}\right)/L}{\operatorname{Tr}\left(\left(\lambda \mathbf{I}+\boldsymbol{\Phi} \boldsymbol{\Pi} \boldsymbol{\Phi}^{\mathrm{T}}\right)^{-1} \mathbf{\Phi}_{\cdot[i]} \mathbf{A}_{i} \mathbf{\Phi}_{\cdot[i]}^{\mathrm{T}}\right)}}
		\label{eq:lr_gamma}
	\end{equation}
	
	{Then, by setting the derivative of $\mathcal{L}(\Theta)$ over $\mathbf{B}$ to zero, we obtain:}
	\begin{equation}
		\widetilde{\mathbf{B}} \leftarrow \sum_{i=1}^{g} \frac{\mathbf{X}_{[i]\cdot}^{\mathrm{T}} \mathbf{A}_{i}^{-1} \mathbf{X}_{[i]\cdot}}{\gamma_{i}}+\eta \mathbf{I}
		\label{eq:lr_Bplus}
	\end{equation}
	\begin{equation}
		\mathbf{B} \leftarrow \frac{\widetilde{\mathbf{B}}}{\left\|\widetilde{\mathbf{B}}\right\|_{\mathcal{F}}}
		\label{eq:lr_B}
	\end{equation}
	
	{Let $\tilde{\mathbf{Y}} \triangleq \mathbf{Y} \mathbf{B}^{-1 / 2}$, $\tilde{\mathbf{X}} \triangleq \mathbf{X} \mathbf{B}^{-1 / 2}$, $\tilde{\mathbf{V}} \triangleq \mathbf{V B}^{-1 / 2}$ to decouple  $\mathbf{B}$ from $\mathbf{A}$ in the original model and then following the EM method\cite{Wan2014IdentifyingTN}, we can derive:}
	\begin{equation}
		\mathbf{A}_{i} \leftarrow \frac{1}{L} \sum_{l=1}^{L} \frac{\tilde{\mathbf{\Sigma}}_{[i]}+\tilde{\boldsymbol{\mu}}_{[i]l} \tilde{\boldsymbol{\mu}}_{[i] l}^{T}}{\gamma_{i}}
		\label{eq:lr_Ai}
	\end{equation}
	where $\widetilde{\boldsymbol{\Sigma}}_{[i]} \in \mathbb{R}^{h \times h}$ is the $i$-th diagonal block of $\tilde{\boldsymbol{\Sigma}}$, and
	\begin{equation}
		\begin{aligned}
			&\widetilde{\boldsymbol{\Sigma}}=\boldsymbol{\Pi}-\boldsymbol{\Pi} \boldsymbol{\Phi}^{T}\left(\lambda \mathbf{I}+ \boldsymbol{\Phi} \boldsymbol{\Pi} \boldsymbol{\Phi}^{\mathrm{T}}\right)^{-1} \boldsymbol{\Phi} \boldsymbol{\Pi}\\
			&\tilde{\boldsymbol{\mu}}=\boldsymbol{\Pi} \boldsymbol{\Phi}^{T}\left(\lambda \mathbf{I}+\boldsymbol{\Phi} \boldsymbol{\Pi} \boldsymbol{\Phi}^{T}\right)^{-1} \mathbf{Y} \mathbf{B}^{-1 / 2}
		\end{aligned}
		\label{eq:lr_hat}
	\end{equation}
	
	{Lastly, the learning rule for $\lambda$ is derived similarly as in \cite{Zhang2011SparseSR} using EM method:}
	\begin{equation}
		\lambda \leftarrow \frac{1}{N L}\|\tilde{\mathbf{Y}}-\boldsymbol{\Phi} \widetilde{\boldsymbol{\mu}}\|_{\mathcal{F}}^{2}+\frac{1}{N} \sum_{i=1}^{g} \operatorname{Tr}\left(\widetilde{\mathbf{\Sigma}}_{[i]} \mathbf{\Phi}_{\cdot[i]}^{\mathrm{T}} \mathbf{\Phi}_{\cdot[i]}\right)
		\label{eq:lr_lambda}
	\end{equation}
	where $\boldsymbol{\Phi}_{\cdot[i]}$ denotes the consecutive columns in $\boldsymbol{\Phi}$ which correspond to the $i$-th block in $\mathbf{X}$, i.e., $\mathbf{X}_{[i]\cdot}$.

	\begin{figure*}[tbp]
		\centering
		\subfigure[{3D human head model and the 12-channel mfEMT coils. The radius of the sensing region is 60 mm.}]{\includegraphics[width =1.6 in]{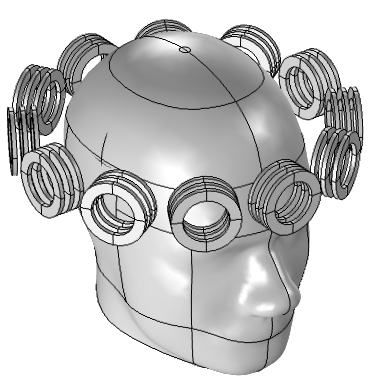}}
		\hspace{0.1in}
		\subfigure[{Magnetic flux (red streamline) and induced current density in the head excited by a coil.}]{\includegraphics[width =1.6 in]{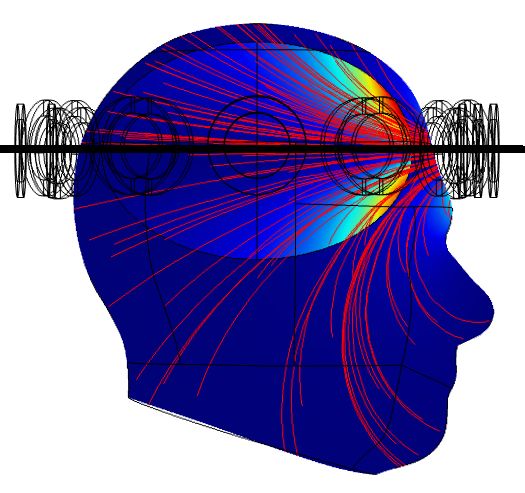}}
		\hspace{0.1in}
		\subfigure[{Phantom 1. $\Omega_0$: air domain; $\Omega_1$: skull; $\Omega_2$: brain; $\Omega_{hem}$: hemorrhagic strokes (radius: 7 mm ).}]{\includegraphics[width =1.6 in]{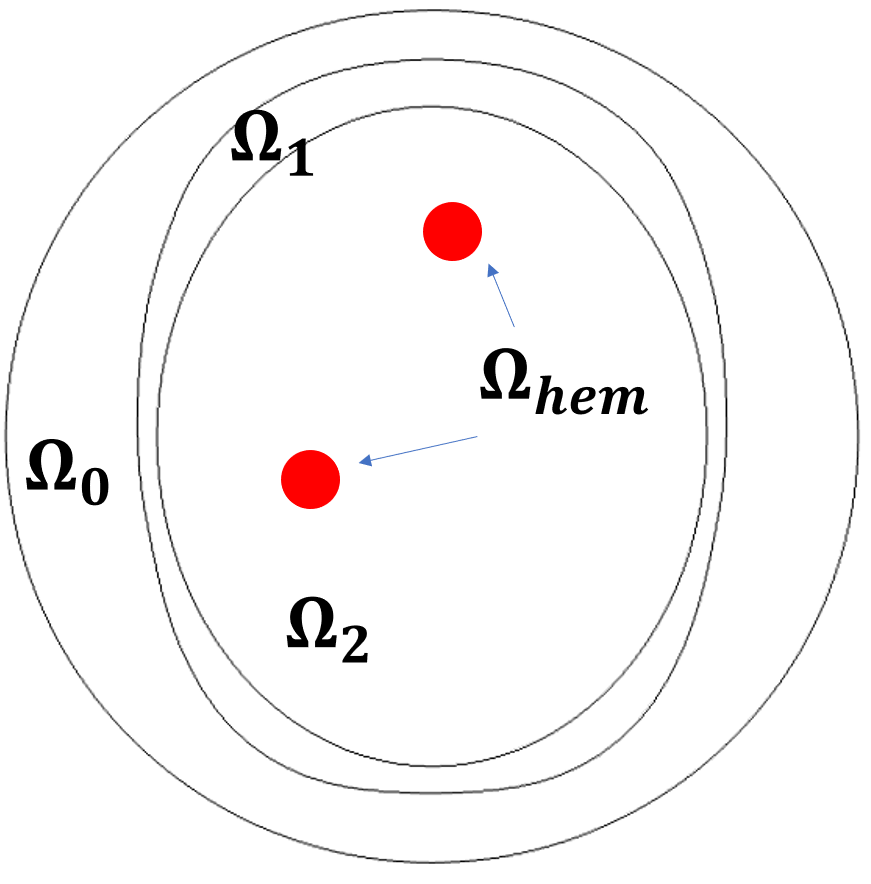}}
		\hspace{0.1in}
		\subfigure[{Phantom 2. $\Omega_0$: air domain; $\Omega_1$: skull; $\Omega_2$: brain; $\Omega_{isc}$: ischaemic strokes (radius: 7 mm).}]{\includegraphics[width = 1.6 in]{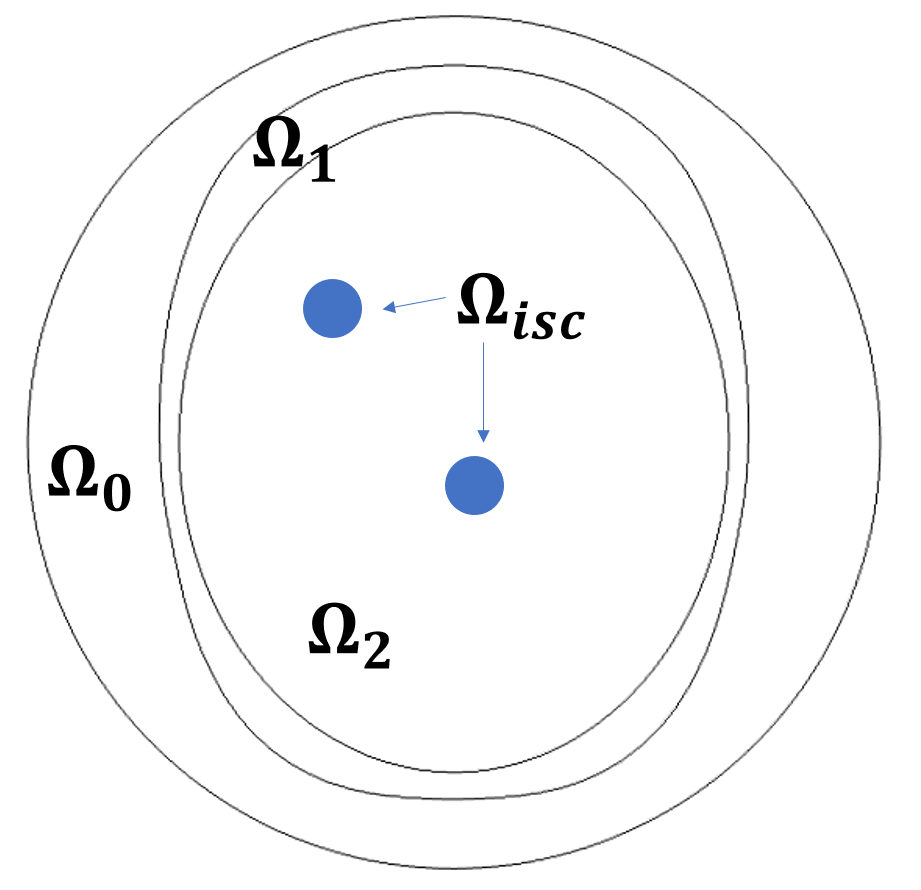}}
		\caption{{Numerical simulation setup of mfEMT for stroke detection. The mfEMT sensor array is established in 3D format shown in (a) but 2D reconstruction is considered on a cut plane (black line) shown in (b); to validate stroke detection, two phantoms in (c) and (d) with different conductivity spectra were created.}}
		\label{fig:head}
	\end{figure*}

	To mitigate the overfitting problem, estimation of $\mathbf{B}$ and $\mathbf{A}_i$ in (\ref{eq:lr_B}), (\ref{eq:lr_Ai}) is further regularized in Toeplitz matrix as formulated in (\ref{eq:toep_times}). Instead of deriving from the cost function, we make empirical estimations of $r_{si}$ and $r_f$:
	\begin{equation}
		\begin{array}{l}
			{\tilde{r}_{f}=\frac{\overline{\operatorname{diag}\left({\mathbf{B}}, 1\right)}}{\overline{\operatorname{diag}\left({\mathbf{B}}\right)}}} \\
			{r_{f}={\operatorname{sign}\left(\tilde{r}_{f}\right) \cdot \min \left\{\left|\tilde{r}_{f}\right|, 0.99\right\}}} 
		\end{array}
		\label{eq:toep_rf}
	\end{equation}
	\begin{equation}
		\begin{array}{l}
			{\tilde{r}_{si}=\frac{\overline{\operatorname{diag}\left({\mathbf{A}_i}, 1\right)}}{\overline{\operatorname{diag}\left({\mathbf{A}_i}\right)}}} \\
			{r_{si}={\operatorname{sign}\left(\tilde{r}_{si}\right) \cdot \min \left\{\left|\tilde{r}_{si}\right|, 0.99\right\}}} 
		\end{array}
		\label{eq:toep_rs}
	\end{equation}
	where  ${\overline{\operatorname{diag}\left({\mathbf{\cdot}}\right)}}$ is the average of the elements along the main diagonal;   ${\overline{\operatorname{diag}\left({\mathbf{\cdot}}, 1\right)}}$ is the average along the main sub-diagonal.

	\textbf{Algorithm 1} summarizes the FC-SBL in pseudo-code.

	\section{Experiments and Results}
	\label{sec:Experiment}
	
	{In this section, we present the results of the proposed FC-SBL method based on numerical simulation and experimental data. A 3D head model was established to mimic real conditions, and in particular, to demonstrate the capability to penetrate the highly resistive skull. Phantom experiments were performed using biological materials. }

	\subsection{Numerical Simulation}
\begin{figure}[tbp]
		\centering
		\includegraphics[width = 3.2in]{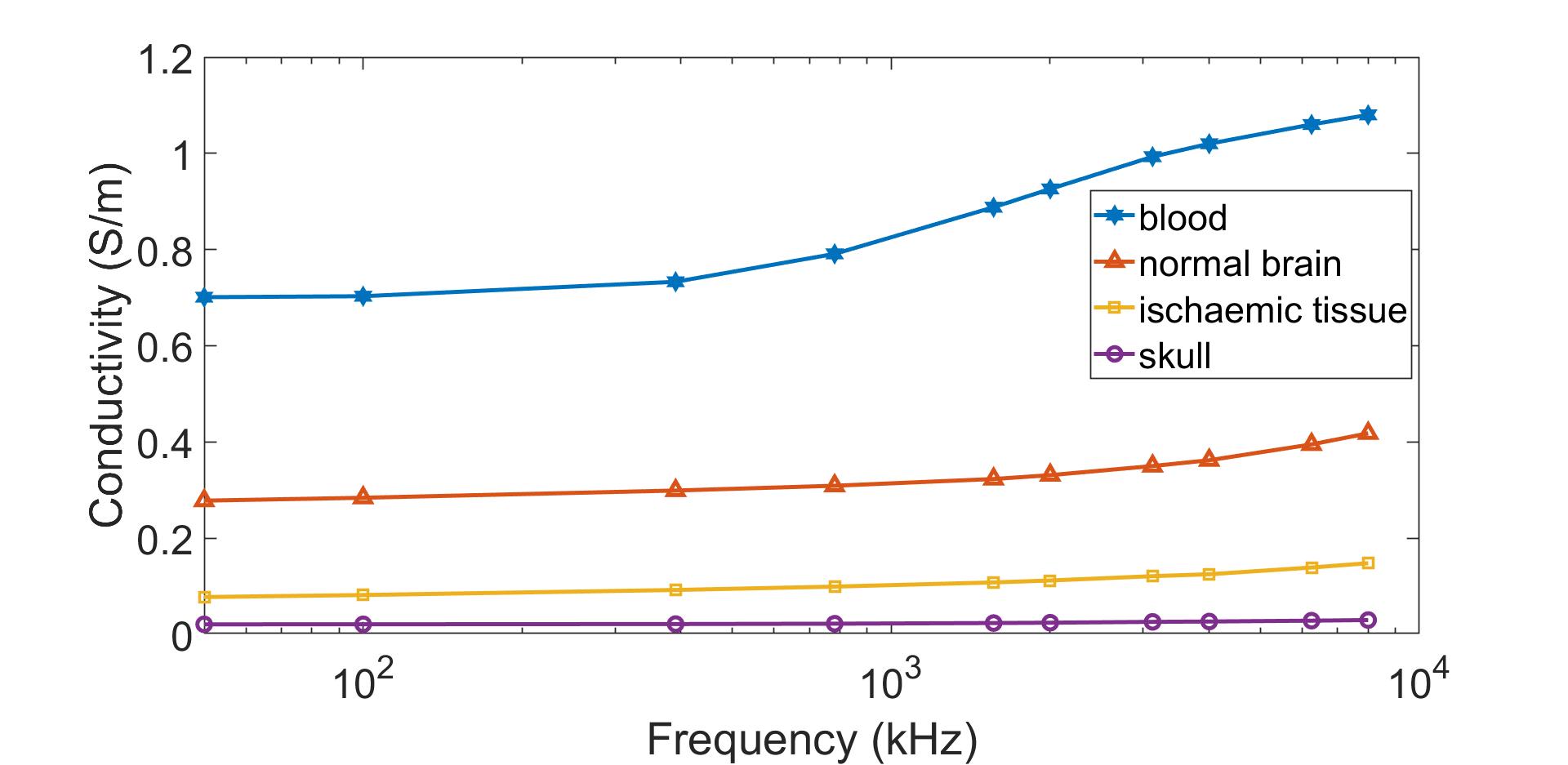}
		\caption{{Conductivity spectra of blood, normal brain tissues, ischemic brain tissues, and skull  {(from 50 kHz to 8 MHz)} \mbox{\cite{gabriel1996compilation, ahn2010frequency}}. }}
		\label{fig:conductivity_head}
	\end{figure}

	{To evaluate the stroke detection performance of mfEMT using FC-SBL, we conducted numerical simulations using COMSOL Multiphysics and established a realistic 3D head model (see Fig. \mbox{\ref{fig:head}}). The 3D head model is the same with SAM phantom provided by IEEE, IEC, and CENELEC from their standard specification of SAR value measurements \cite{ieee2003ieee}. The model was created using free tetrahedral mesh with approximately 80,000 elements. The radius of the sensing area is 60 mm (see Fig. \ref{fig:head} (a) and (b)), with the 12-channel mfEMT coils placed around its periphery. In this model, the skull, brain, stroke anomaly, and mfEMT sensors are simulated. The model is simplified to represent the brain with ellipsoid and the stroke with cylinders. Two cylindrical inclusions of radius 7 mm and height 20 mm were placed in the brain, where Fig. \ref{fig:head} (c) and (d) simulates the hemorrhagic and ischemic stroke, respectively. Conductivity properties of brain tissues and strokes were obtained from \mbox{\cite{gabriel1996compilation, ahn2010frequency}} (see Fig. \mbox{\ref{fig:conductivity_head}}).	Hemorrhagic stroke is caused by a rupture of a blood vessel in the brain whereas ischemic stroke is caused by a block in blood flow \mbox{\cite{horesh2006some}}. 	Fig. \ref{fig:conductivity_head} shows clear distinctions among three conductivity values $(\sigma_{\text{blood}} > \sigma_{\text{brain}} > \sigma_{\text{ischaemia}})$, reinforcing that bio-impedance could distinguish ischaemic stroke from haemorrhagic stroke.	}
	
	{Before applying the background subtraction described in Section \mbox{\ref{sec:preprocess}}, the geometry of the head is extracted first because the effective sensing area is the head domain ($\Omega_1, \Omega_2, \Omega_{hem} /\Omega_{isc}$) excluding the air domain $\Omega_0$. In realistic scenarios, the geometry of the head can be estimated by some positioning apparatuses ahead of diagnosis. It is pointed out in\mbox{\cite{jang2015detection}} that the background of the head domain is highly heterogeneous due to the presence of skull, which is one major reason of the failure of the weighted-difference EIT method based on homogeneous background assumption. On the contrary, mfEMT is not sensitive to the skull and consequently, the influence of such heterogeneity is negligible. Note that we made some simplifications of the model. For instance, we didn't consider the gray matter, white matter, and other tissues in the brain separately. For a highly heterogeneous background of brain tissues, more advanced background subtractions such as non-linear fraction method \mbox{\cite{malone2014stroke}}, or multiple weighted-difference method \mbox{\cite{jang2015detection}} can be adopted. However, in this paper we primarily focus on demonstrating mfEMT as a potential imaging modality for stroke detection and FC-SBL as an effective image reconstruction method for this application. }

	\begin{table*}[tbp]
		\caption{{ {Image reconstruction results based on simulation data ($\text{SNR}_{f_1} = 30\ dB$)}}}
		\begin{spacing}{1}
			\centering
			\begin{tabular}{m{0.5cm}m{2.2cm}m{2.2cm}m{2.2cm}|m{2.2cm}m{2.2cm}m{2.2cm}}
				& \begin{tabular}[c]{@{}l@{}}TV  \cite{borsic2009vivo}\\ (hemorrhagic)  \end{tabular}  
				& \begin{tabular}[c]{@{}l@{}}SA-SBL \cite{Liu2018ImageRI}\\ (hemorrhagic)  \end{tabular} 
				& \begin{tabular}[c]{@{}l@{}}FC-SBL\\ (hemorrhagic)  \end{tabular} 
				& \begin{tabular}[c]{@{}l@{}}TV \cite{borsic2009vivo}\\ (ischemic)   \end{tabular} 
				& \begin{tabular}[c]{@{}l@{}}SA-SBL \cite{Liu2018ImageRI}\\ (ischemic)    \end{tabular} 
				& \begin{tabular}[c]{@{}l@{}}FC-SBL\\ (ischemic)    \end{tabular}  \\
				$f_1$ & 
				\includegraphics[width=2cm]{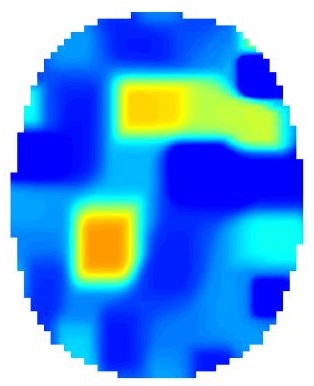} & 
				\includegraphics[width=2cm]{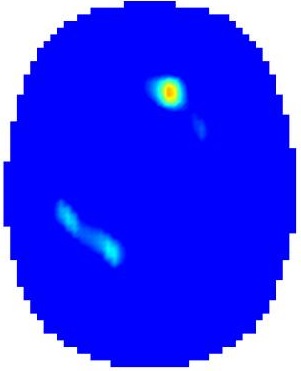} &
				\includegraphics[width=2cm]{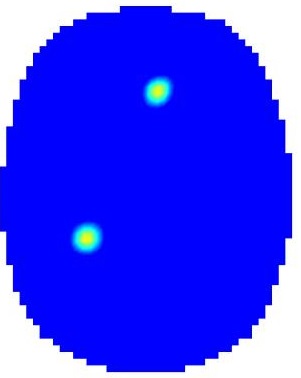} &
				\includegraphics[width=2cm]{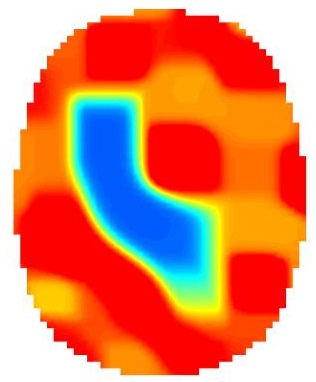} & 
				\includegraphics[width=2cm]{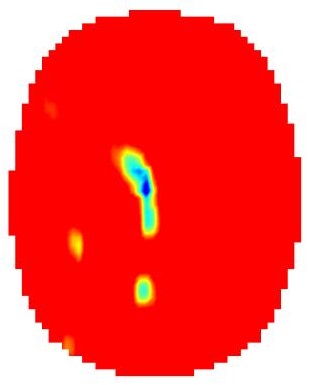} & 
				\includegraphics[width=2cm]{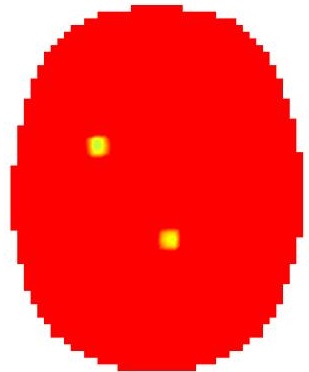}  \\ 
				$f_2$ & 
				\includegraphics[width=2cm]{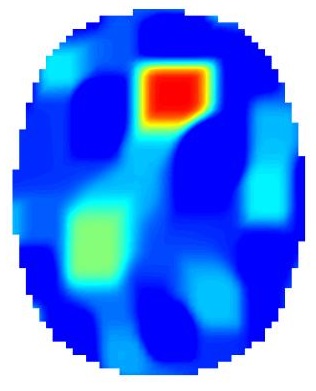} & 
				\includegraphics[width=2cm]{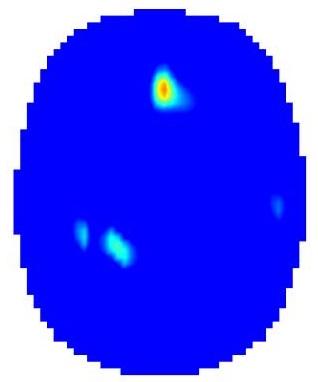}	&
				\includegraphics[width=2cm]{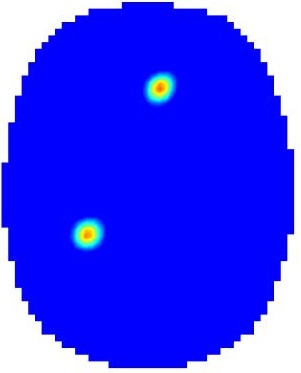}	&
				\includegraphics[width=2cm]{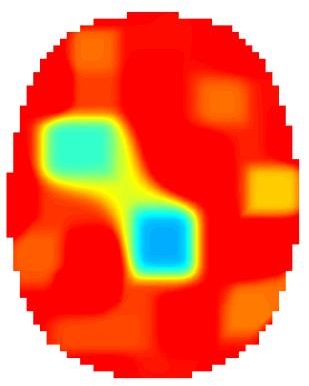} & 
				\includegraphics[width=2cm]{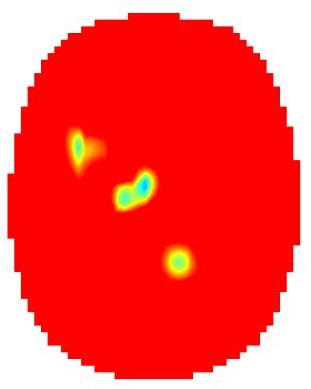} & 
				\includegraphics[width=2cm]{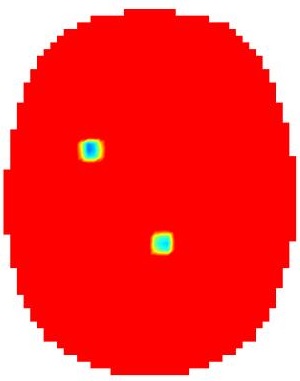}  \\
				$f_3$ & 
				\includegraphics[width=2cm]{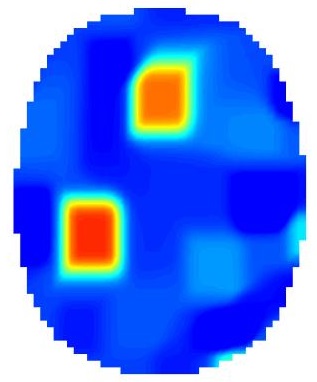} & 
				\includegraphics[width=2cm]{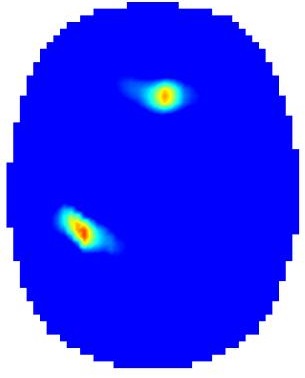}	&
				\includegraphics[width=2cm]{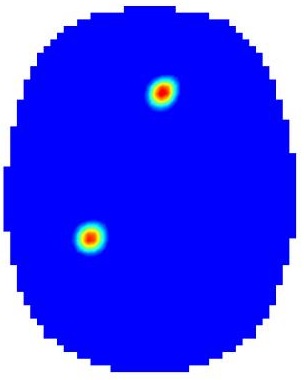}	&
				\includegraphics[width=2cm]{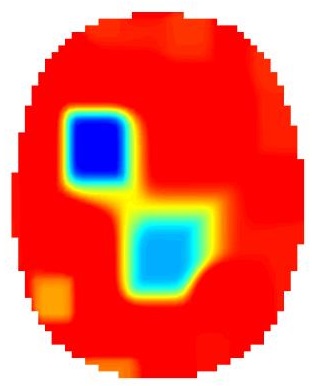} & 
				\includegraphics[width=2cm]{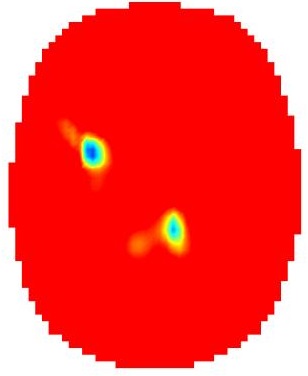} & 
				\includegraphics[width=2cm]{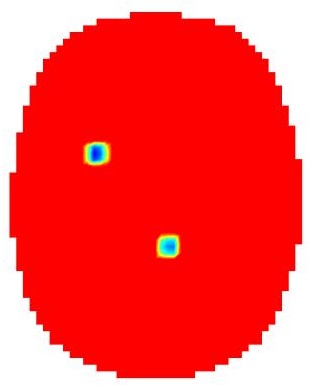}  \\
				$f_4$ & 
				\includegraphics[width=2cm]{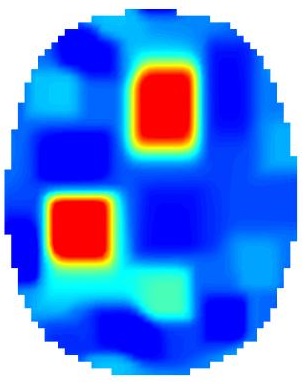} & 
				\includegraphics[width=2cm]{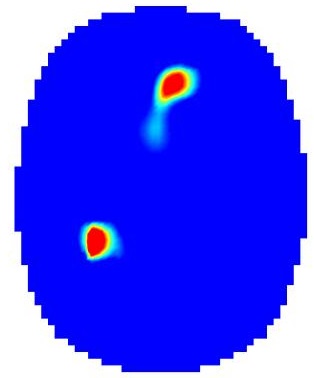}	&
				\includegraphics[width=2cm]{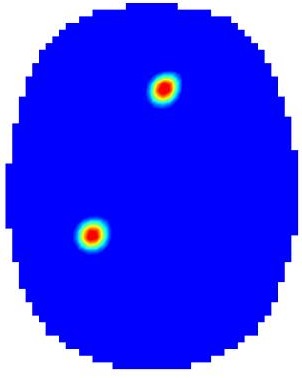}	&
				\includegraphics[width=2cm]{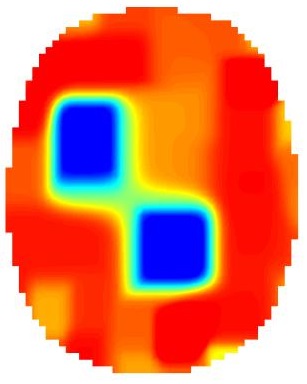} & 
				\includegraphics[width=2cm]{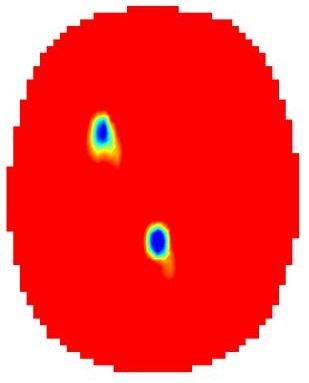} & 
				\includegraphics[width=2cm]{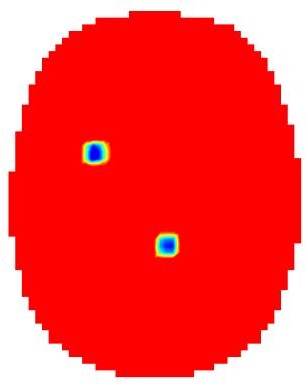}  \\
				& \includegraphics[width=2.2cm]{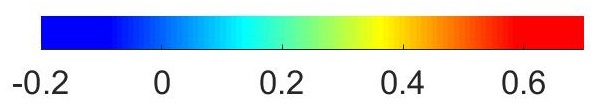}
				& \includegraphics[width=2.2cm]{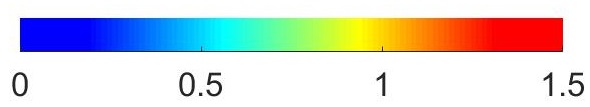}
				& \includegraphics[width=2.2cm]{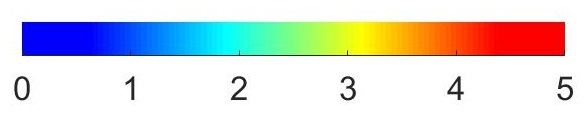}
				& \includegraphics[width=2.2cm]{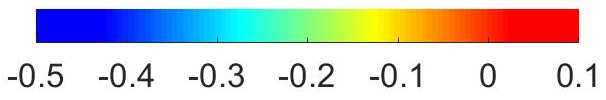}
				& \includegraphics[width=2.2cm]{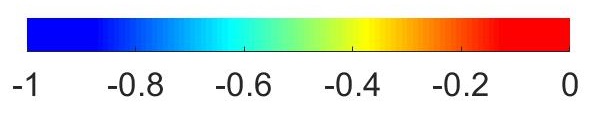}
				& \includegraphics[width=2.2cm]{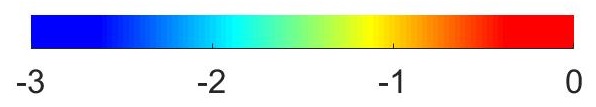}\\
			\end{tabular}
		\end{spacing}
		\label{tab:FCSBL_phantom}
	\end{table*}
	
	\begin{figure}[tbp]
		\centering
		\includegraphics[width = 3.2in]{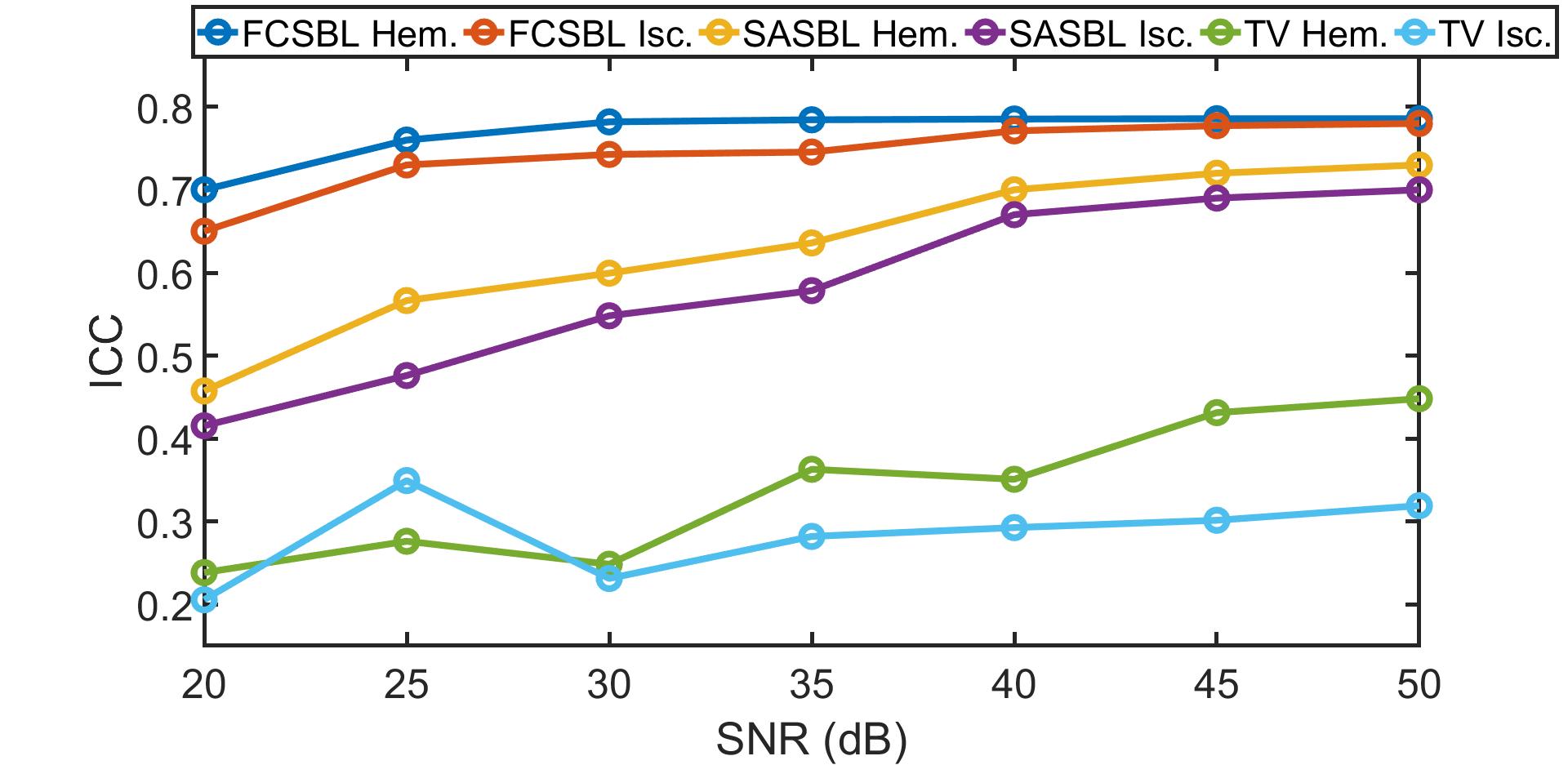}
		\caption{{Performance comparison in terms of ICC.}}
		\label{fig:icc}
	\end{figure}
	
	\begin{figure}[tbp]
		\centering
		\includegraphics[width = 3.2in]{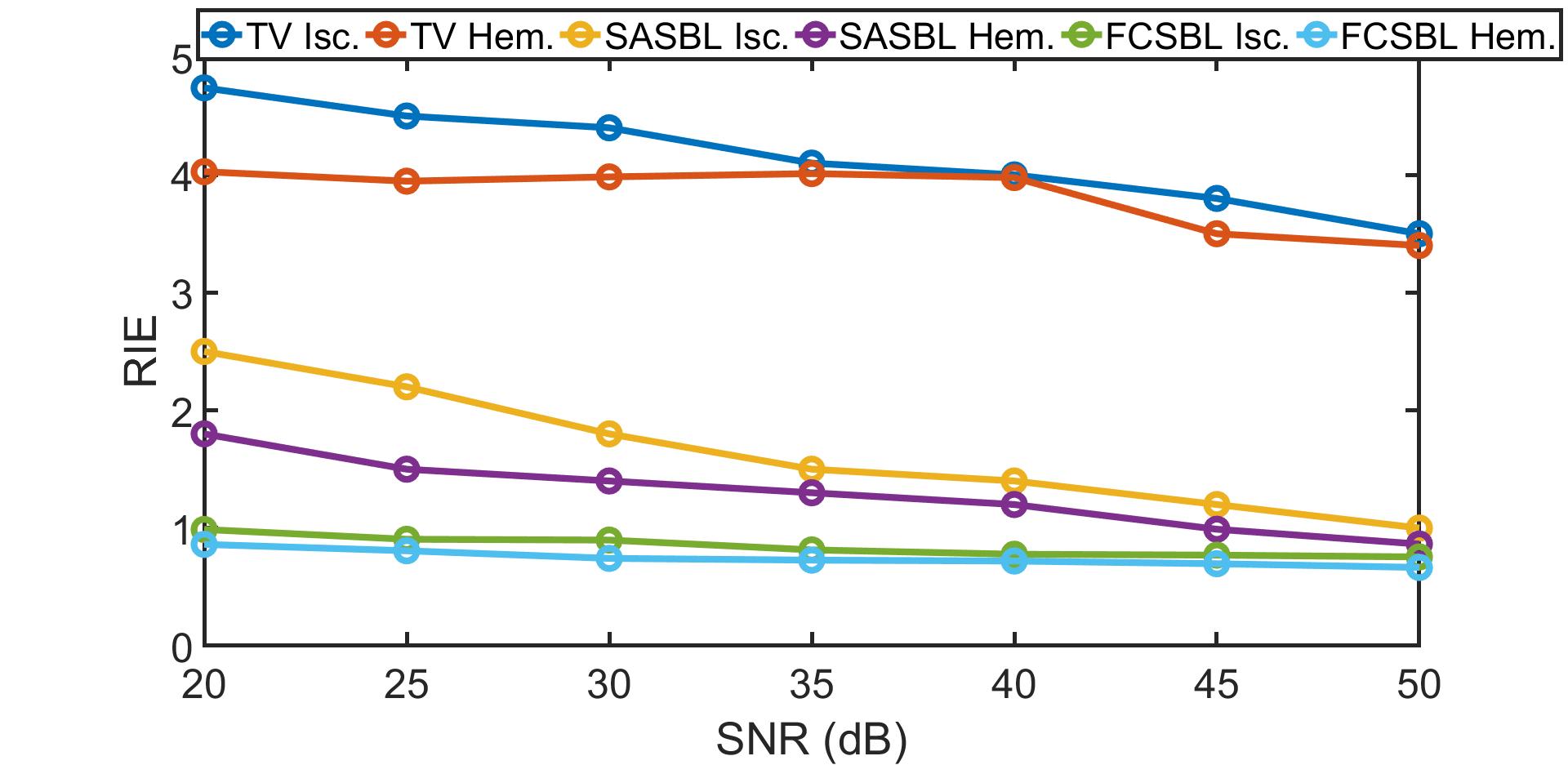}
		\caption{{Performance comparison in terms of RIE.}}
		\label{fig:rie}
	\end{figure}

	{Five excitation frequencies $[f_0, f_1,f_2,f_3,f_4] = [1/16, 1/8, 1/4, 1/2, 1]\times 6.25\ MHz$ are selected to fulfil full-cycle sampling in order to avoid spectral leakage in FFT, given that the sampling rate of the system is $62.5\ MHz$.  The reference frequency for FD imaging is $f_0$. In order to simulate the background noise of the real system, we add the same quantity of white noise to each measurement. As the conductivity changes with respect to $f_0$ increase with frequency, the resulting SNR also increases correspondingly. The noise power is determined to ensure $\text{SNR}_{f_1}=30\ dB$.}
	
	{Table \mbox{\ref{tab:FCSBL_phantom}} illustrates the image reconstruction results of the simulated phantom. The first three columns show reconstruction images of phantom 1 (hemorrhagic case) by using Total Variation (implemented with primal dual inner point method) \mbox{\cite{borsic2009vivo}}, SA-SBL\mbox{\cite{Liu2018ImageRI}}, and FC-SBL, respectively, while rows correspond to four different frequencies from $f_1$ to $f_4$.  
	The regularization coefficient of TV is $\alpha_{TV}=2\times 10^{-6}$ and termination conditions are $\epsilon_{\mathrm{min}}  = 1\times 10^{-5}$ and $\vartheta_{\mathrm{max}} = 20$. The termination conditions of SA-SBL and FC-SBL are set as $\epsilon_{\mathrm{min}}  = 1\times 10^{-5}$ and $\vartheta_{\mathrm{max}} = 120$.  The 2D block size of FC-SBL is  $h=9$.
	Computations were carried out for four times to solve each measurement for TV and SA-SBL, whereas FC-SBL solves multiple measurements simultaneously. The results demonstrate that in the first column, there are strong artifacts in the images and the shapes of objects are hardly visible. As for the results of SA-SBL in the second column, image quality is greatly improved with fewer artifact and distortion. Also note that more artifact and distortions can be observed at lower frequencies, i.e. $f_1$ and $f_2$. This is consistent with the conclusion from\mbox{\cite{Liu2018ImageRI}} that the performance of SA-SBL will degrade at low SNR (e.g. 35 dB or less).  In comparison, results based on FC-SBL (see column 3) better preserves the shape and location of stroke phantom for all frequencies. The underlying reason is that FC-SBL exploits the frequency constraints among multiple measurements and the underlying structural pattern could be better recovered even under strong noise levels. 
	Phantom 2 (ischemic case) results in the last three columns show negative conductivity changes with the increase of frequency. This case is more challenging because the conductivity changes of the ischemic case with frequency is much smaller than those of phantom 1 (see Fig. \ref{fig:conductivity_head}). We can see that both TV and SA-SBL fail to differentiate the two anomalies at $f_1$ and $f_2$. In contrast, FC-SBL apparently outperforms SA-SBL and TV by recovering the stroke pattern under all frequencies more accurately.}

	To quantitatively evaluate the performance of algorithms, we adopt two widely used criteria, i.e. Image Correlation Coefficient (ICC) and Relative Image Error (RIE):
	\begin{equation}
		\operatorname{ICC}=\frac{\sum_{i=1}^{M}\left(\sigma_{i}-\bar{\sigma}\right)\left(s_{i}-\bar{s}\right)}{\sqrt{\sum_{i=1}^{M}\left(\sigma_{i}-\bar{\sigma}\right)^{2} \sum_{i=1}^{M}\left(s_{i}-\bar{s}\right)^{2}}}
		\label{eq:ICC}
	\end{equation}
	
	\begin{equation}
		\operatorname{RIE}=\frac{\|\sigma-s\|_{2}}{\left\|_{S}\right\|_{2}}
		\label{eq:RIE}
	\end{equation}
	where $s$ is the ground-truth conductivity; $\sigma$ is the reconstructed conductivity.

	{We evaluated the reconstructed images at $f_1$ under different SNRs ranging from 20 dB to 50 dB. At each SNR, we conducted 1000 repeated experiments of each algorithm. Fig. \mbox{\ref{fig:icc}} and \mbox{\ref{fig:rie}} show the comparison results in terms of ICC and RIE. It can be observed that FC-SBL outperforms SA-SBL for both ICC and RIE, especially when the SNR is low. When the SNR approaches 45 dB, SA-SBL and FC-SBL show comparable performance. In comparison, TV regularization yields the largest RIE and the smallest ICC, indicating the worst results.}

	\subsection{System Evaluation}
	
\begin{figure}[tbp]
		\centering
		\includegraphics[width = 3.2 in]{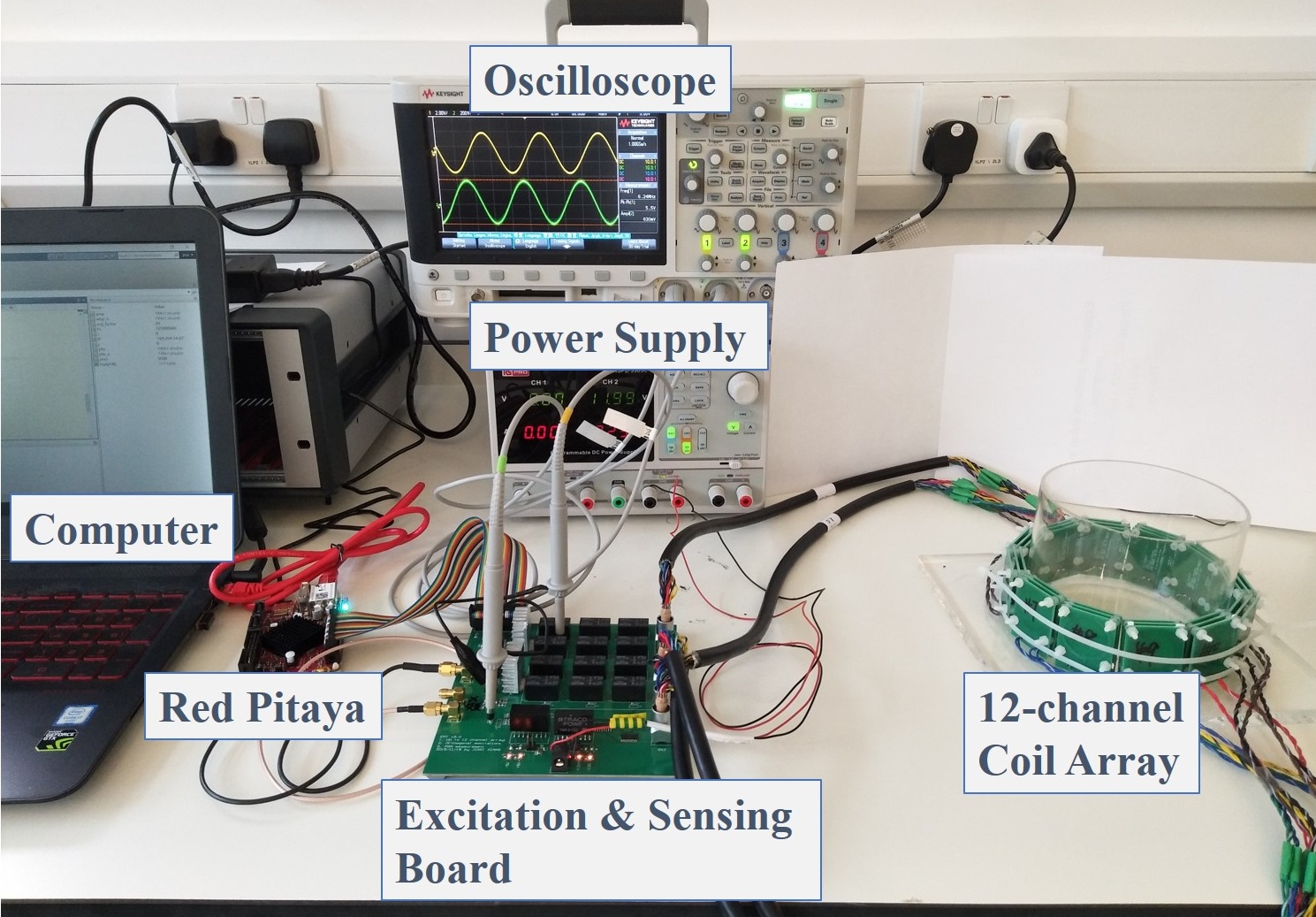}
		\caption{Experimental setup of the 12-channel mfEMT system.}
		\label{fig:System2}
	\end{figure}

	Fig. \ref{fig:System2} shows the experimental setup of the mfEMT system. It consists of a 12-channel coil array with a diameter of 120 mm, a signal generation and data acquisition module based on Red Pitaya, a computer for image reconstruction, and a circuit board that incorporates multiplexer, excitation and sensing electronics.
	
\begin{figure}[tbp]
		\centering
		\includegraphics[width = 3.4in]{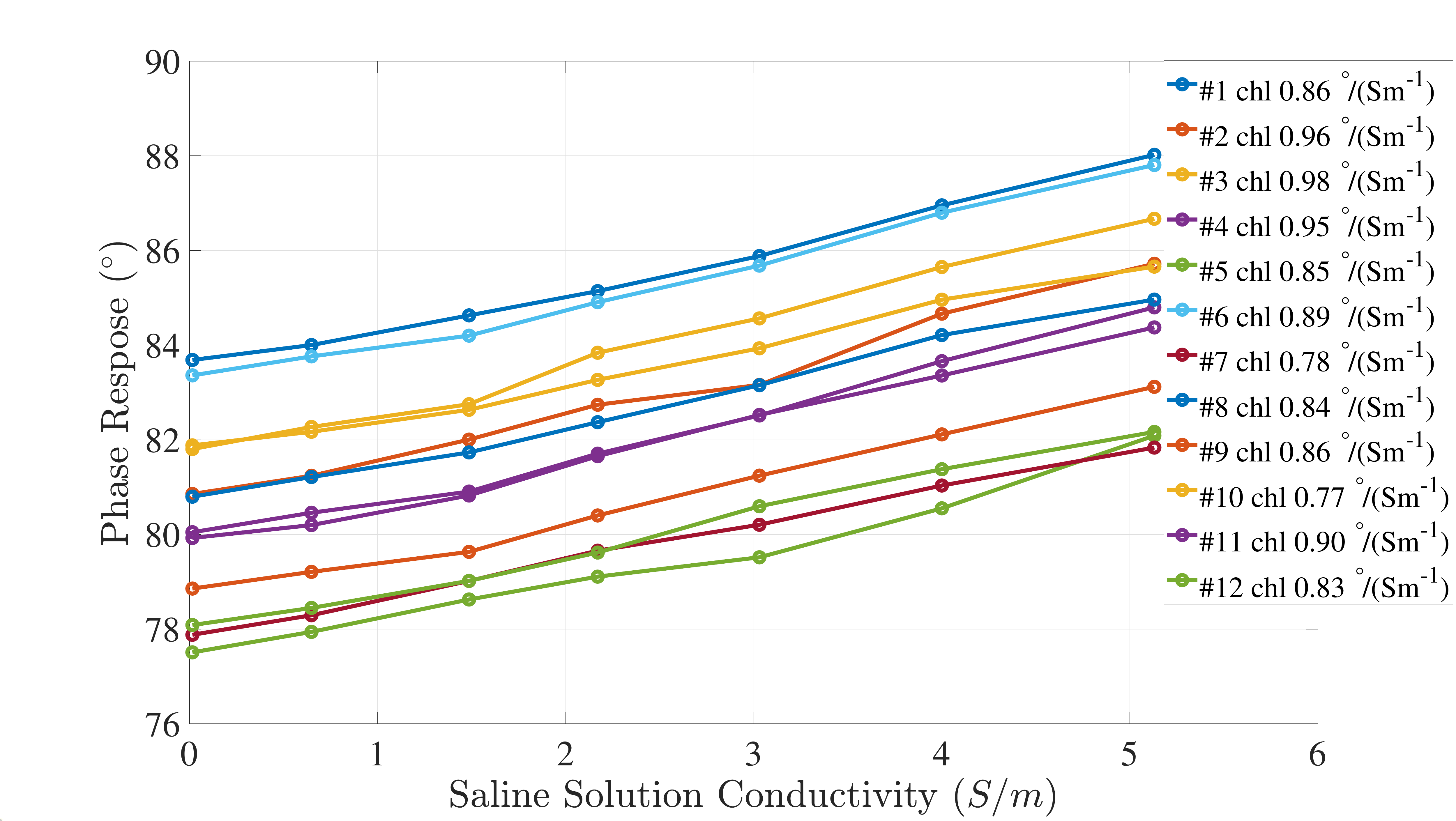}
		\caption{Sensitivity calibration of 12 channel sensors.}
		\label{fig:SensCal}
	\end{figure}

	{We conduct the sensitivity calibration of 12 channels in order to compensate the manufacturing inconsistency of sensor coils.} The sensitivity is calibrated using a sequence of NaCl solutions with conductivity ranging from 0.01 S/m to 5.13 S/m. The NaCl solution in plastic bottles is placed closely to the coil, which is driven by a 6.25 MHz sinusoidal signal. As shown in Fig. \ref{fig:SensCal}, the  sensitivity of the gradiometer coils ranges from  $0.77\ ^{\circ}/(S\cdot m^{-1})$ to  $0.98\ ^{\circ}/(S\cdot m^{-1})$. The measured phase response is highly linear with conductivity. The difference among each sensor coil is then compensated in the system.

	\subsection{Phantom Experiments}
\begin{figure}[tbp]
		\centering
		\subfigure[ {Phantom 3}, object: cucumber.]{\includegraphics[width =1.5 in]{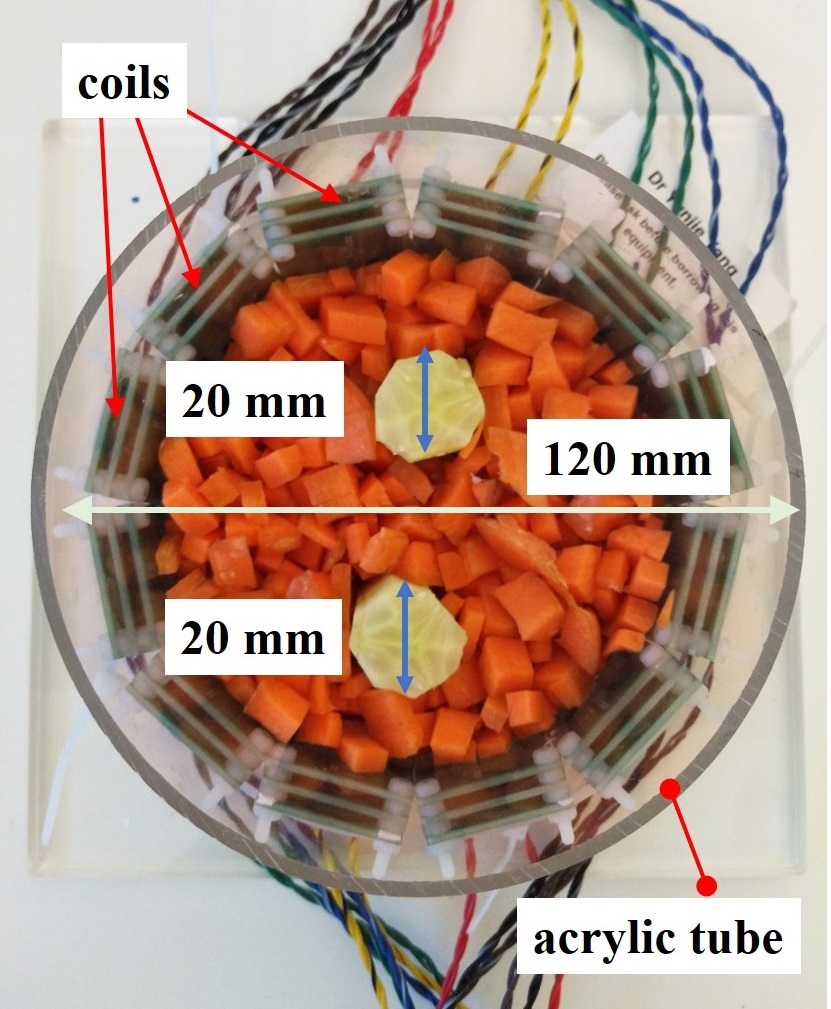}}
		\hspace{0.1in}
		\subfigure[ {Phantom 4}, object: banana.]{\includegraphics[width =1.5 in]{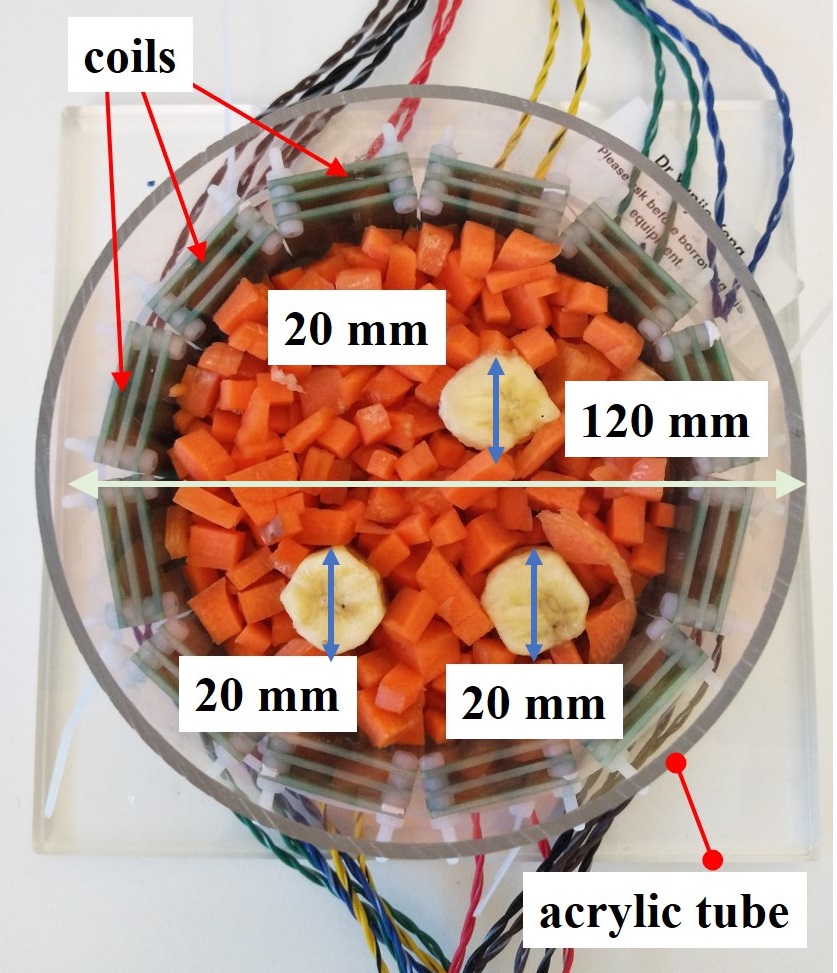}}
		\caption{{Experimental phantoms consist of carrot, banana, and cucumber. Both phantoms use (carrot + saline) as background substance, and a 3-mm thick acrylic wall simulates the effect of skull isolating the biological materials and sensors.}}
		\label{fig:Phantom}
	\end{figure}

	We acquired experimental data using the 12-channel mfEMT system. Two phantoms (see Fig. \ref{fig:Phantom}) were designed using {a 3-mm thick acrylic cylindrical tank, which simulates the effect of skull. There is no direct electrical contact between sensor coils and biological materials.} The tank was filled with background objects, which is a mixture of 0.9\%  sodium chloride solution and small pieces of carrot. Three banana cylinders with a diameter of approximately 20 mm and two cucumber cylinders with a diameter of about 20 mm were placed in the background to simulate the abnormalities in the human brain.  The excitation frequencies in the experiment are $[f_0, f_1,f_2,f_3,f_4] = [1/16, 1/8, 1/4, 1/2, 1]\times 6.25\ MHz$.
	
	\begin{table*}[htb]
		\centering
		\caption{Image reconstruction results based on experimental data.}
		\begin{spacing}{1.5}
			\centering
			\begin{tabular}{m{2cm}m{3cm}m{3cm}m{3cm}m{3cm}m{0.5cm}} 
				& $f_1= 0.7813\ MHz$ & $f_2=1.5625\ MHz$ & $f_3=3.125\ MHz$ & $f_4=6.25\ MHz$ & \\ 
				SA-SBL  {Phantom 3}  & 
				\includegraphics[height=1.2 in]{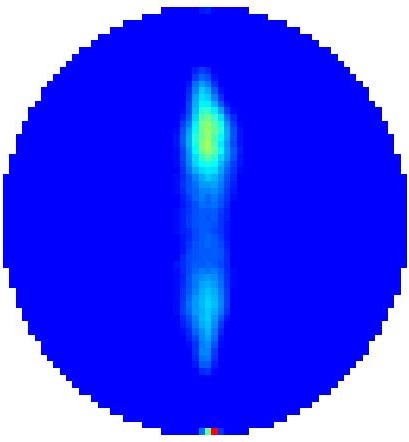} & 
				\includegraphics[height=1.2 in]{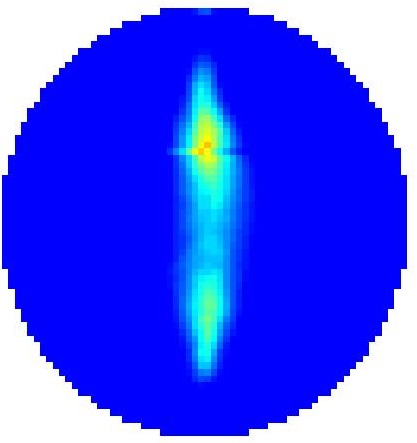} & 
				\includegraphics[height=1.2 in]{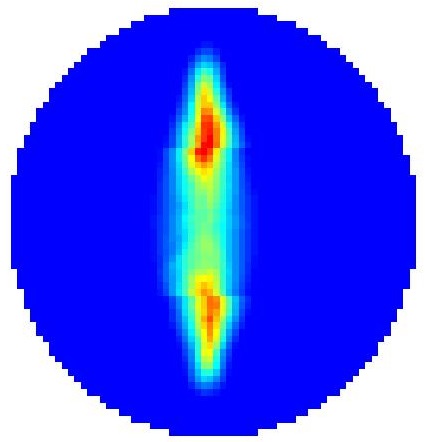} & 
				\includegraphics[height=1.28 in]{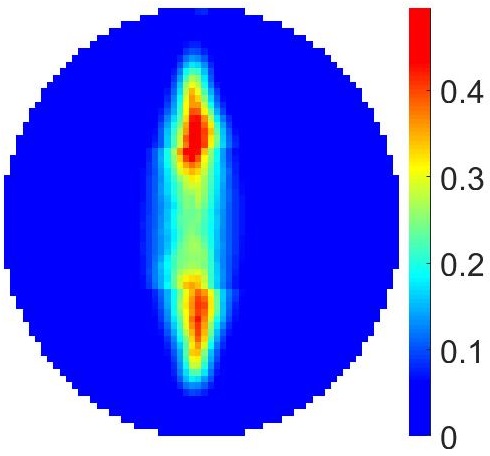} & \\ 
				FC-SBL  {Phantom 3} & 
				\includegraphics[height=1.2 in]{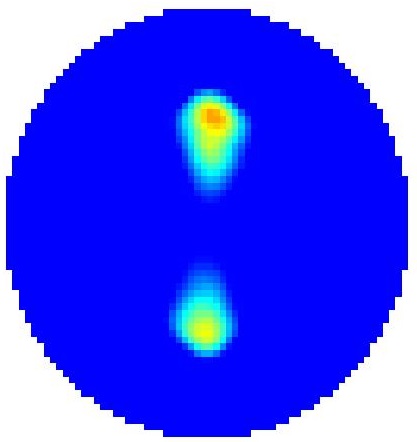} & 
				\includegraphics[height=1.2 in]{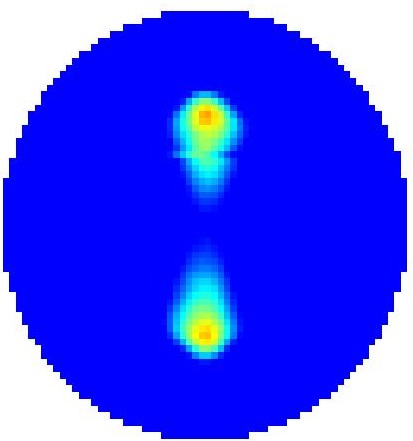} & 
				\includegraphics[height=1.2 in]{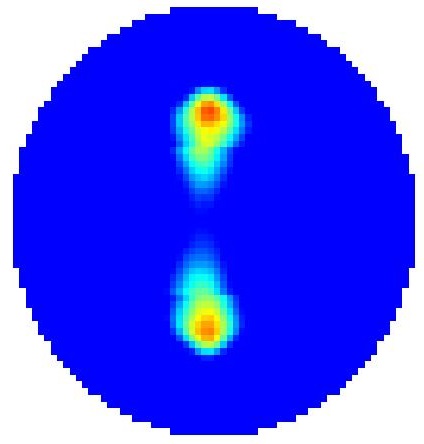} & 
				\includegraphics[height=1.28 in]{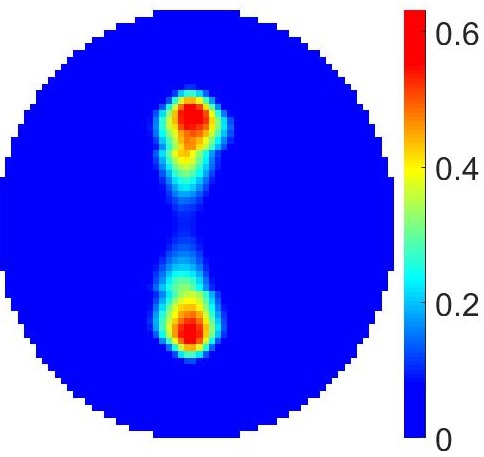} & \\
				SA-SBL  {Phantom 4} & 
				\includegraphics[height=1.2 in]{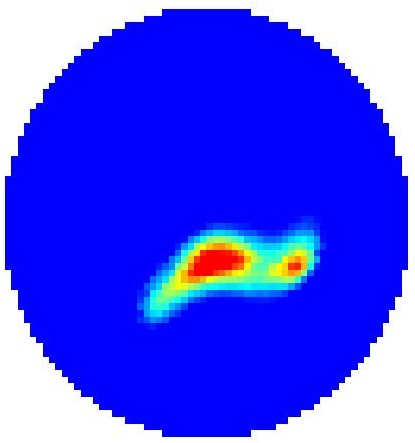} & 
				\includegraphics[height=1.2 in]{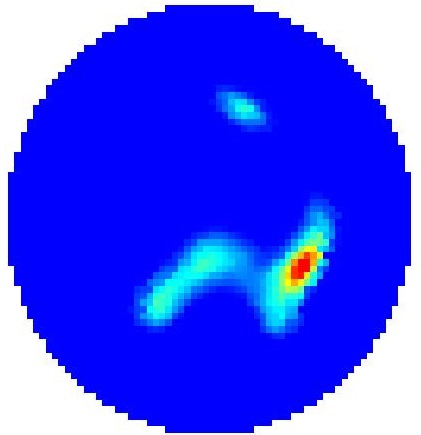} & 
				\includegraphics[height=1.2 in]{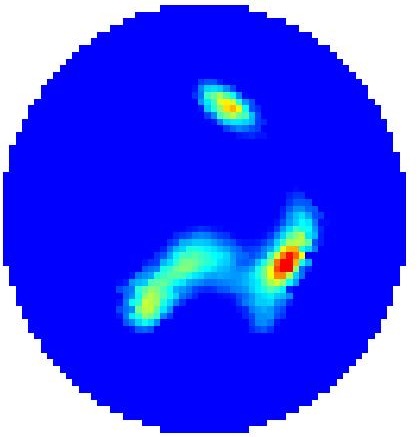} & 
				\includegraphics[height=1.28 in]{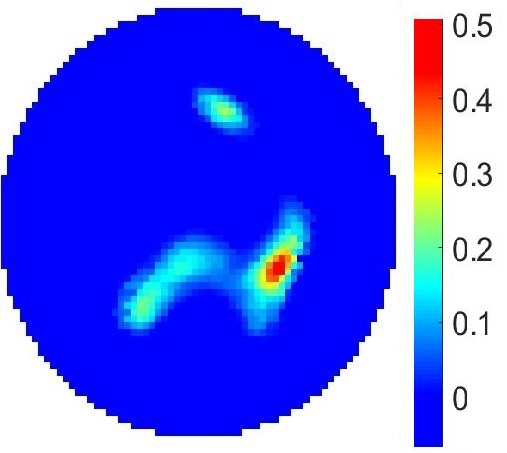} & \\
				FC-SBL  {Phantom 4} & 
				\includegraphics[height=1.2 in]{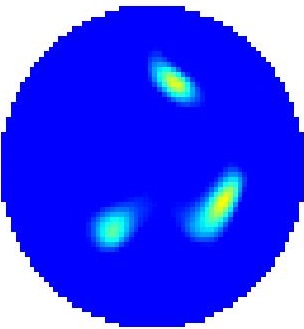} & 
				\includegraphics[height=1.2 in]{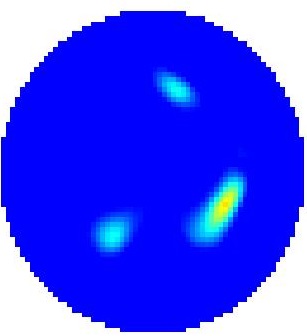} & 
				\includegraphics[height=1.2 in]{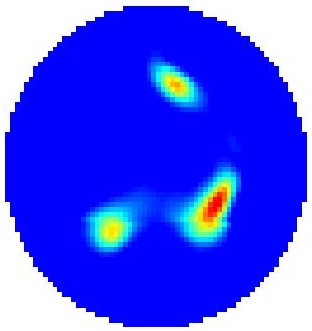} & 
				\includegraphics[height=1.28 in]{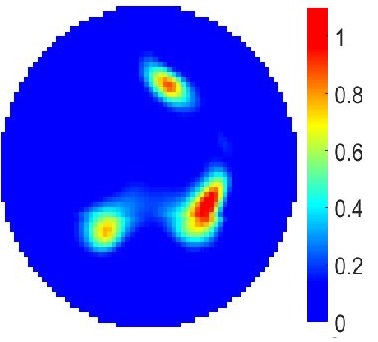} & \\
			\end{tabular}
		\end{spacing}
		\label{tab:Experiment}
	\end{table*}

	Table \ref{tab:Experiment} shows {cross-sectional 2D images} reconstruction results of two phantoms using experimental data. 
	{In experiments, we compare the performance of the proposed FC-SBL with recently reported SA-SBL \cite{Liu2018ImageRI}.} 
	The termination conditions of SA-SBL and FC-SBL are set as $\epsilon_{\mathrm{min}}  = 1\times 10^{-5}$ and $\vartheta_{\mathrm{max}} = 120$.  The 2D block size of FC-SBL is  $h=9$ (length of 2D block is 3).  Other parameters are determined as default in the paper.
	
	Overall, for both methods, the image quality increases with frequency, which is reasonable as the conductivity of biological materials is increasing monotonically, which leads to a larger signal response. For  {phantom 3}, SA-SBL recovers the two target objects and their location, whilst there are some obvious artifacts between two objects. By comparing, FC-SBL generates more accurate images with clear boundaries and less artifacts. With respect to   {phantom 4}, the superiority of FC-SBL over SA-SBL is more distinct.  {phantom 4}  is more challenging to reconstruct than  {phantom 3}, because the inter-tissue inductive coupling issue \cite{xiao2017effect} of multiple objects might lead to more severe nonlinear effect.  {Another reason is  the reduced conductivity contrast between cucumber and carrot compared to that of banana and carrot in phantom 3.} We can observe that the performance of SA-SBL for  {phantom 4} degrades considerably, especially at lower frequencies, i.e. $f_1$ and $f_2$. Differently, FC-SBL recovers the three objects well throughout the whole frequency range. Although there exists a little shape deformation, each individual object and its location are clearly resolved.

	{
		\section{Discussion}
		\label{sec:discussion}
		In this section, we discuss some practical issues regarding the algorithm  development of FC-SBL.
		\subsection{Embedding Block Size}
		As stated in Section \ref{sec:FCSBL_body}, we use overlapped 2D blocks to represent the unknown objects. The assumption of equal block size $h$ actually is not critical for image resolution in practical use. When the size of an object is larger than $h$, it can be recovered by a set of overlapped blocks and if the size of an object is smaller than $h$, it can be recovered by tuning a smaller spatial correlation coefficient $r_{si}$. This tuning process is automatically learned. Specifically, Eq. (\ref{eq:lr_gamma}) determines the block sparsity pattern, and Eq. (\ref{eq:lr_Ai}) parameterizes the spatial correlation within blocks.
		\subsection{Noise Assumption}
		In FC-SBL, we assume the noise vector $ \mathbf{v} _i $ is Gaussian and its sources are mutually independent. It is in accordance with the fact that in experiments, the thermal noise of the electronics follows Gaussian distribution\cite{whitaker2005electronics}. The system noise, such as the internal noise from excitation modules or motion artifacts caused by patients, is not considered in the algorithm. The system noise in experimental observations usually comes from the sensing instruments and is difficult to be encapsulated in the algorithm. System errors other than random noise could possibly be removed at the data preprocessing stage by means of sensitivity calibration, normalization, etc., ahead of applying the image reconstruction algorithm.
		\subsection{Computation Time}
		This work adopts the linearized mfEMT model, therefore it is not required to update the forward model during image reconstruction as the non-linear reconstruction method \cite{malone2015reconstruction} does. Hence, its implementation is much less time-consuming. When implementing FC-SBL, the maximum iteration number was 120 and the elapsed time per iteration for simultaneously reconstructing four images was 10.516s on a PC with MATLAB 2019b, 32GB RAM memory and a 6-core Intel i7-8700 CPU. In comparison, SA-SBL took 8.886s per each iteration for one image on the same computation platform, i.e. equivalently 35.544s for four images. FC-SBL is more time-efficient than SA-SBL for multiple measurements, but larger RAM memory is required for FC-SBL because the MMV model in (\ref{eq:MMV3}) learns the parameters in a higher dimensional space ($N\times gh$ system matrix) compared to the single vector problems ($N\times M$ system matrix) in (\ref{eq:SMV}). 
	}
	
	\section{Conclusion}
	\label{sec:conclusion}
	To tackle the challenge of early diagnosis of acute strokes, this paper presented a mfEMT system to perform bioimpedance imaging in a non-radiative, noncontact, and {baseline-free} manner, making it promising for applications  where it is not possible to obtain a 'baseline' measurement under healthy condition. We introduced a Sparse Bayesian Learning (SBL) approach for image reconstruction of mfEMT using the frequency constraints on multi-frequency measurements. The proposed FC-SBL framework provides a Bayesian inversion approach that has plug-and-play specifications of the forward model and strong regularizing and parameter-free properties for inverse problems with both spatial and frequency correlations among measurements. {The  proposed FC-SBL method was validated through simulations and experiments, showing the benefits of FC-SBL in stroke imaging with improved image resolution and robustness to noise.}
	
	{Future work will focus on
		(1) investigating and mitigating the effects of inhomogeneous backgrounds, mismatch between sensors and model, and realistic noise from patient recordings as these factors are the likely limitations to hinder mfEMT for clinical use \cite{ljungqvist2017clinical, goren2018multi}; 
		(2) extending the experimental setup to assess the imaging performance of the dynamic evolution of acute stroke, and 
		(3) extending the proposed FC-SBL method for 3D image reconstruction, considering the brain as inherently 3D geometries \cite{Liu2019AcceleratedSS}. }

	
	%

	%



	\ifCLASSOPTIONcaptionsoff
	\newpage
	\fi

	\bibliography{IEEEabrv,mybiblio}
	
	%
	
	%
	%
	%
	
	
	

\end{document}